\documentclass[useAMS,usenatbib]{mn2e}
\usepackage{graphicx}
\usepackage{times}

\begin{document}

\newcommand{\bfit}[1]{\mbox{\boldmath $#1$}}
\newcommand{\bfsf}[1]{\mbox{\boldmath \sf #1}}

\title[Accurate  seeing  measurements  with  MASS  and  DIMM]{Accurate
seeing measurements with MASS and DIMM}

\author[Tokovinin \& Kornilov]{A.~Tokovinin$^{1}$\thanks{E-mail:
atokovinin@ctio.noao.edu},
V.~Kornilov$^2$\thanks{E-mail:
victor@sai.msu.ru} \\
$^1$Cerro Tololo Inter-American Observatory, Casilla 603, La Serena, Chile\\
$^2$Sternberg Astronomical Institute, Universitetsky prosp. 13, 119992 Moscow, Russia
}

\date{-}

\pagerange{\pageref{firstpage}--\pageref{lastpage}} \pubyear{2007}

\maketitle

\label{firstpage}

\begin{abstract}

Astronomical seeing  is quantified  by a single  parameter, turbulence
integral, in  the framework of  the Kolmogorov turbulence  model. This
parameter  can be routinely  measured by  a Differential  Image Motion
Monitor, DIMM.  A  new instrument, Multi-Aperture Scintillation Sensor
(MASS), permits  to measure  the seeing in  the free  atmosphere above
$\sim$0.5km and,  together with a  DIMM, to estimate  the ground-layer
seeing.  The absolute  accuracy of both methods is  studied here using
analytical   theory,  numerical   simulation,   and  experiments.    A
modification of  the MASS data processing to  compensate for partially
saturated scintillation  is developed.  We  find that the DIMM  can be
severely   biased   by  optical   aberrations   (e.g.   defocus)   and
propagation.   Seeing  measurements  with  DIMM  and  MASS  can  reach
absolute  accuracy  of  $\sim$10\%  when their  biases  are  carefully
controlled.  Pushing this limit to 1\% appears unrealistic because the
seeing itself is just  a model-dependent parameter of a non-stationary
random process.
\end{abstract}


%----------------------------------------------------------------------
\section{Introduction}

Measurements  of astronomical ``seeing''  are performed  for selecting
new sites and supporting operation of existing telescopes. Recently, a
standard   Differential    Image   Motion   Monitor    (DIMM)   method
\citep{Martin87,DIMM}  has  been complemented  with  a new  technique,
Multi-Aperture  Scintillation Sensor  (MASS)  \citep{MASS}.  This  new
instrument is  based on the  analysis of scintillation and  permits to
measure  the seeing  in the  free atmosphere,  isoplanatic  angle, and
adaptive-optics (AO) time constant.  Both MASS and DIMM require only a
small  telescope   and  can  be  combined  in   a  single  instrument
\citep{Kor07}.   MASS-DIMM  site   monitors  gradually  become  a  new
standard.

We can  evaluate the  ground-layer (GL) seeing  produced in  the first
0.5\,km above the observatory  by subtracting the turbulence integrals
measured with DIMM and MASS.  It is important to measure the GL seeing
for evaluating the performance of Ground-Layer Adaptive Optics (GLAO).
However, subtraction  only works when both methods  are accurate, i.e.
deliver results  on the absolute  scale.  Absolute accuracy  of seeing
measurements  becomes  critical for  the  site  comparison, where  the
differences are often below 10\%.

In principle,  both DIMM  and MASS should  give accurate  results when
their  instrument parameters  are  set correctly  and their  intrinsic
biases are understood and removed. Here we investigate these biases in
detail, quantify  them, and propose  corrections.  Biases of  the DIMM
method    have   been    already   addressed    in    the   literature
\citep{Martin87,PASP02}.  We continue by considering additional effects
such as  light propagation and  optical aberrations.  The  analysis of
the MASS  method given  by \citet{Rest} is  extended by  studying small
departures from the weak-scintillation  theory which cause a systematic
bias,  ``over-shoots'',  and  can  be  corrected by  a  modified  data
processing.

This work has been stimulated by  the need to get accurate results from
the  existing  suite  of   MASS-DIMM  instruments,  described  in  the
accompanying  paper  \citep{Kor07}. Correct  setting  and operation  of
these  instruments,  also  critical  for  getting  accurate  data,  is
addressed  in that  paper.   Here we  concentrate  on the  theoretical
analysis    and   simulations,    trying   to    formulate   practical
recommendations and recipes  while avoiding mathematical complexity as
much as possible.

We begin  by asking the question  ``what is seeing  and how accurately
can  it  be  defined  and  measured?''   in  Sect.~\ref{sec:def}.   In
Sect.~\ref{sec:MASS}, the  MASS method  is studied under  conditions of
realistic (not  weak) scintillation.  Then  in Sect.~\ref{sec:DIMM} we
address two previously neglected effects  in a DIMM -- propagation and
optical  aberrations --  and show  that they  can cause  a significant
bias,  in addition  to the  known  DIMM biases.   Our conclusions  and
recommendations are formulated in Sect.~\ref{sec:concl}.

\section{What is measured by a seeing monitor?}
\label{sec:def}

%-----------------------------------------------------
\subsection{Turbulence parameters}

 Turbulence   measurements   are   based   on  the   standard   theory
\citep{Tatarskii,Roddier81}.    For  convenience,   major  atmospheric
parameters are recalled  in Table~\ref{tab:seeing}.  These definitions
assume  a  scale-free Kolmogorov  turbulence  spectrum  with only  one
parameter,  turbulence   strength.   This  single   parameter  can  be
expressed  equivalently  by  $\varepsilon_0$,  $r_0$, or  $J$.   These
quantities  (except $J$)  depend on  the imaging  wavelength $\lambda$
which  is assumed  here  to  be 500\,nm  if  not specified.   Accurate
measurement  of this single  parameter (called  ``seeing'' in  a broad
sense)  is the purpose  of a  seeing monitor.   The atmosphere  can be
split into an arbitrary number of  zones (or layers), and we also want
to measure the seeing produced by each of the layers -- the turbulence
profile.   Other atmospheric  parameters  (time constant,  isoplanatic
angle), not considered here, are also of interest to modern astronomy.

\begin{table}
\center
\caption{Quantities relevant to seeing}
\label{tab:seeing}
\medskip
\begin{tabular}{l l l }
\hline
\hline
Quantity & Units & Formula \\
\hline
Seeing (FWHM) & rad & $\varepsilon_0 = 0.98 \lambda/r_0$ \\
%= (J/6.8\,10^{-13})^{3/5}$ \\
Turbulence integral & m$^{1/3}$ & $J = \int_{\rm path} C_n^2(z)\, {\rm  d}z$ \\ 
Fried parameter & m & $r_0  = [0.423 (2 \pi / \lambda)^2 J]^{-3/5}$ \\
 & & $ r_0 = 1.01 \lambda/\varepsilon_0$ \\
Phase power spectrum & m$^2$ & $\Phi_\varphi({\bfit f})  =   C \; | {\bfit f}|^{-11/3}$, \\
  & &   $ C  =    0.00969 (2 \pi/ \lambda)^2 J $ \\
& &  $C = 0.0229 r_0^{-5/3}$ \\
\hline
\hline
\end{tabular}
\end{table}

The seeing $\varepsilon_0$ is often considered to be the angular image
spread  caused  by  the   atmosphere.   This  interpretation  is  only
approximate   because  even   in  a   perfect  telescope   the  actual
long-exposure   point  spread  function   (PSF)  depends   on  several
additional  parameters (outer  scale, wavelength,  telescope diameter,
guiding).    Current   seeing  monitors   measure   only  one   number
$\varepsilon_0$  which  is necessary,
but not sufficient  for accurate prediction of the  atmospheric PSF or
other turbulence-related optical quantities.

It  is always implicitly  assumed that  the statistical  properties of
turbulence are stationary, while in fact they are not. Any measurement
refers only  to the  particular moment in  time and to  the particular
viewing  direction.   The   non-stationarity  precludes  very  precise
measurements of atmospheric parameters because averaging over a larger
spatial or temporal sample of turbulence does not lead to the improved
statistical  precision.   This aspect,  often  overlooked, also  makes
seeing measurements  intrinsically irreproducible. Comparisons between
seeing monitors or between a monitor  and a telescope can be only made
in  a  statistical  sense,  with associated  non-stationarity  errors.
Fortunately, in  a MASS-DIMM instrument both channels  sample the same
turbulent path,  hence non-stationarity does not  affect the precision
of the GL seeing estimate.

The  seeing is  not a  well-defined physical  quantity like  length or
mass,  so  the  intrinsic   accuracy  of  seeing  monitors  cannot  be
arbitrarily  high.  It is  unrealistic to  expect a  relative accuracy
better than 1\%  because the ``seeing'' cannot be  defined with such a
high  accuracy.  It  is shown  below  that keeping  biases within  few
percent is not easy.

%-----------------------------------------------------
\subsection{Seeing monitors}

\begin{figure}
\centerline{\includegraphics[width=8cm]{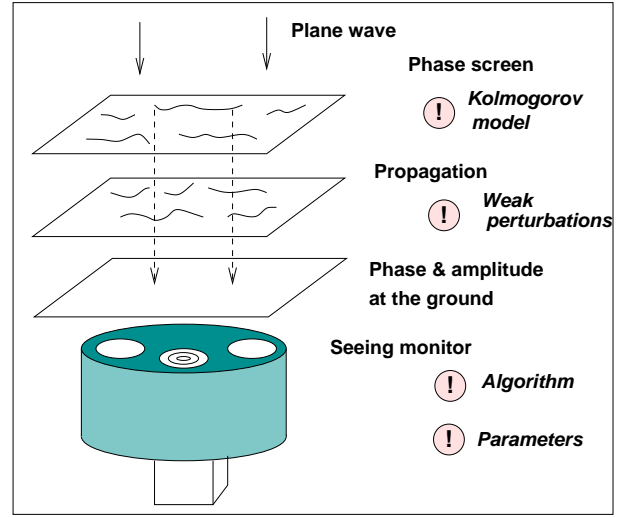} }
\caption{Schematic   representation   of   a  seeing   monitor.    The
exclamation  marks   and  text  in  italics   show  approximations  or
uncertainties present in any seeing measurements.
\label{fig:seeingmon} }
\end{figure}

A seeing monitor measures  some statistical properties of phase and/or
amplitude of a  light wave at the ground and  interprets them in terms
of the model parameter, seeing. Several approximations are involved in
this process, as illustrated in Fig.~\ref{fig:seeingmon} and discussed
below. 

Both DIMM  and MASS  are sensitive to  phase distortions  with spatial
scales below 1\,m where the  Kolmogorov model works well.  A DIMM with
small apertures is  not affected by the finite  turbulence outer scale
$L_0$  \citep[less than 1\% bias on variance for
$L_0  > 4$\,m, cf.][]{Borgnino92}.   On the  other hand,  the absolute
image  motion in a  small 10-cm  telescope is  influenced by  a finite
$L_0$, typically at $\sim$10\% level.  Thus, a site-testing instrument
based on  the absolute (non-differential) image motion  gives a biased
$r_0$  estimate if  $L_0$  is  not known.   In  practise, such  seeing
monitors  are  no  longer  used  because they  are  also  affected  by
mechanical errors  (wind shake, tracking).  For the  same reasons, the
theoretically  perfect interferometric  method of  seeing measurements
\citep{Interf} has not become widely adopted.

%-----------------------------------------------------
\subsection{Propagation}
\label{sec:prop}

Seeing measurements are affected by the light propagation transforming
pure  phase   distortions  to  a   mixture  of  phase   and  amplitude
distributions.  The  spatial spectra of the light  phase $\varphi$ and
logarithm of  amplitude $\chi$ after passing through  a weak turbulent
layer and propagation over a distance $z$ are
\begin{eqnarray}
 \Phi_\varphi({\bfit f}) & = &  0.0229 \; r_0^{-5/3} | {\bfit f}|^{-11/3}
 \cos^2(\pi \lambda z  | {\bfit f}|^2 ) , 
\label{eq:Wphi} \\
 \Phi_\chi({\bfit f}) & = &  0.0229 \; r_0^{-5/3} | {\bfit f}|^{-11/3}
 \sin^2(\pi \lambda z  | {\bfit f}|^2 ) .
\label{eq:Wchi}
\end{eqnarray}
The spectrum of intensity  fluctuations (scintillation) is $\Phi_I = 4
\Phi_\chi$. The  amplitude and phase  are not correlated at  any given
point, but their cross-spectrum is not zero at ${\bfit f} \neq 0$, being
proportional to the product of the sine and cosine terms.

The {\it scintillation index} $s^2$ is defined as
\begin{equation}
s^2 = \langle \Delta I^2  \rangle   / \langle I \rangle ^2     ,
\label{eq:sig_I}
\end{equation}
where $I$  is the instantaneous light intensity  received through some
aperture,  $\Delta I$  is its  fluctuation. In  the small-perturbation
regime,  $s^2 \ll 1$,  formula (\ref{eq:sig_I})  is equivalent  to the
variance  of the  $\log I  =  2 \chi  $.  The  scintillation index  is
calculated then by  integrating the amplitude spectrum (\ref{eq:Wchi})
with  a suitable  aperture filter.   The effect  of  several turbulent
layers is simply additive.

\begin{figure}
\centerline{\includegraphics[width=4cm]{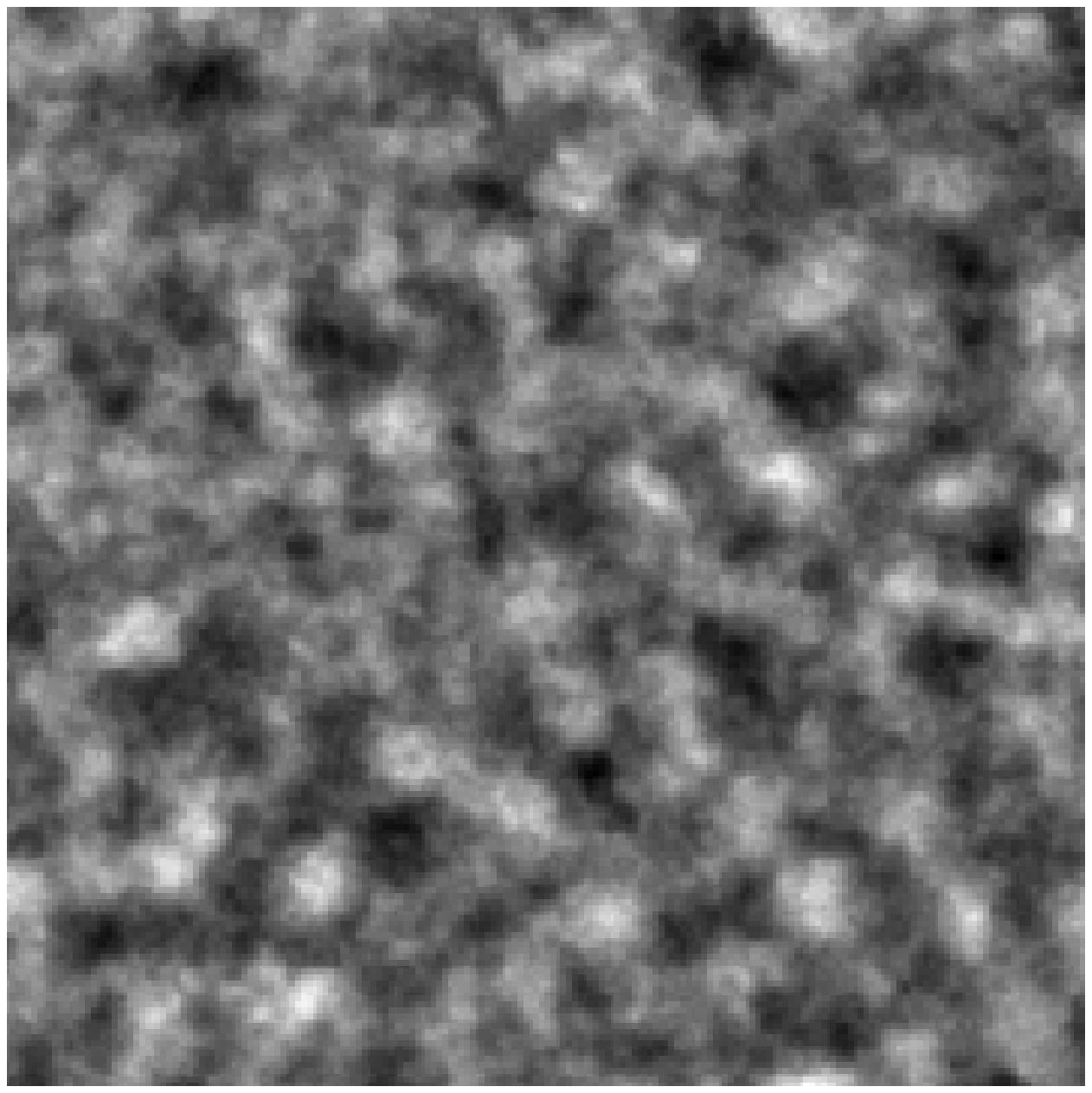} 
\includegraphics[width=4cm]{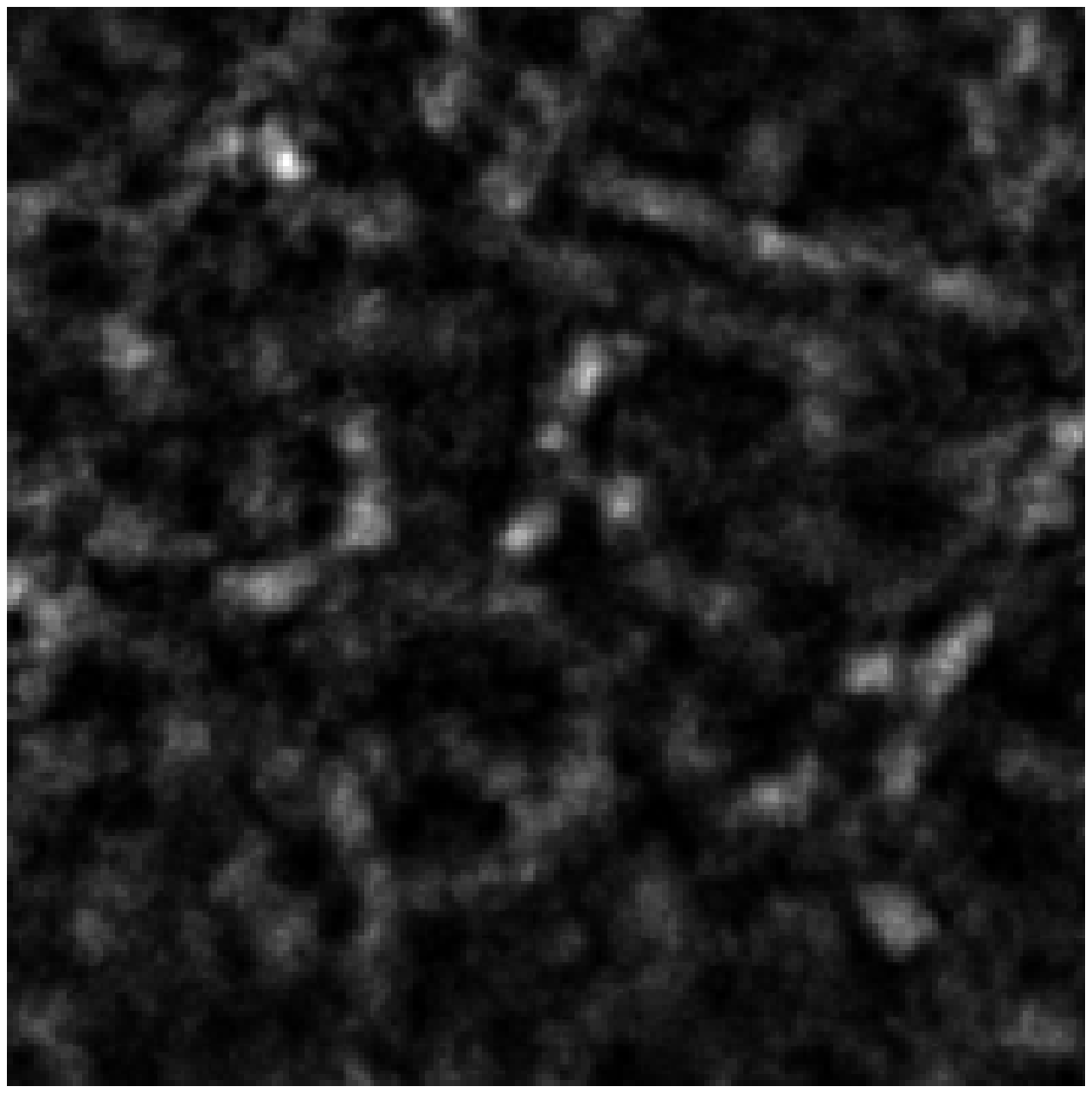} }
\caption{Intensity  screens  of  1~m$^2$  size produced  by  a  single
 turbulent  layer  and  10\,km   propagation  in  conditions  of  weak
 scintillation (left, 0\farcs2 seeing) and strong scintillation (right,
 1$''$    seeing).   Wavelength    0.45\,$\mu$m.   
\label{fig:intens} }
\end{figure}

\begin{figure}
  \includegraphics[width=8cm]{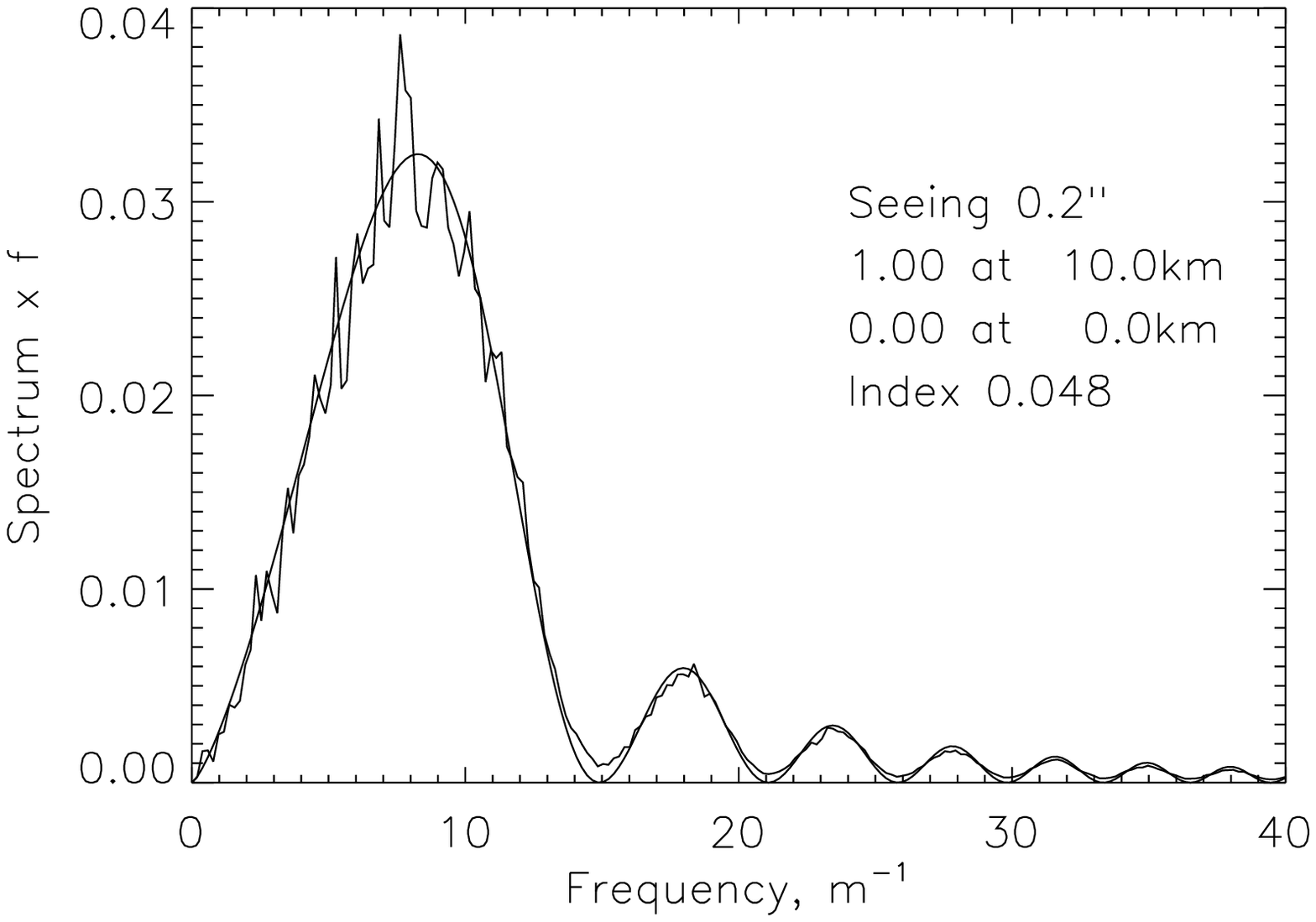} 
\includegraphics[width=8cm]{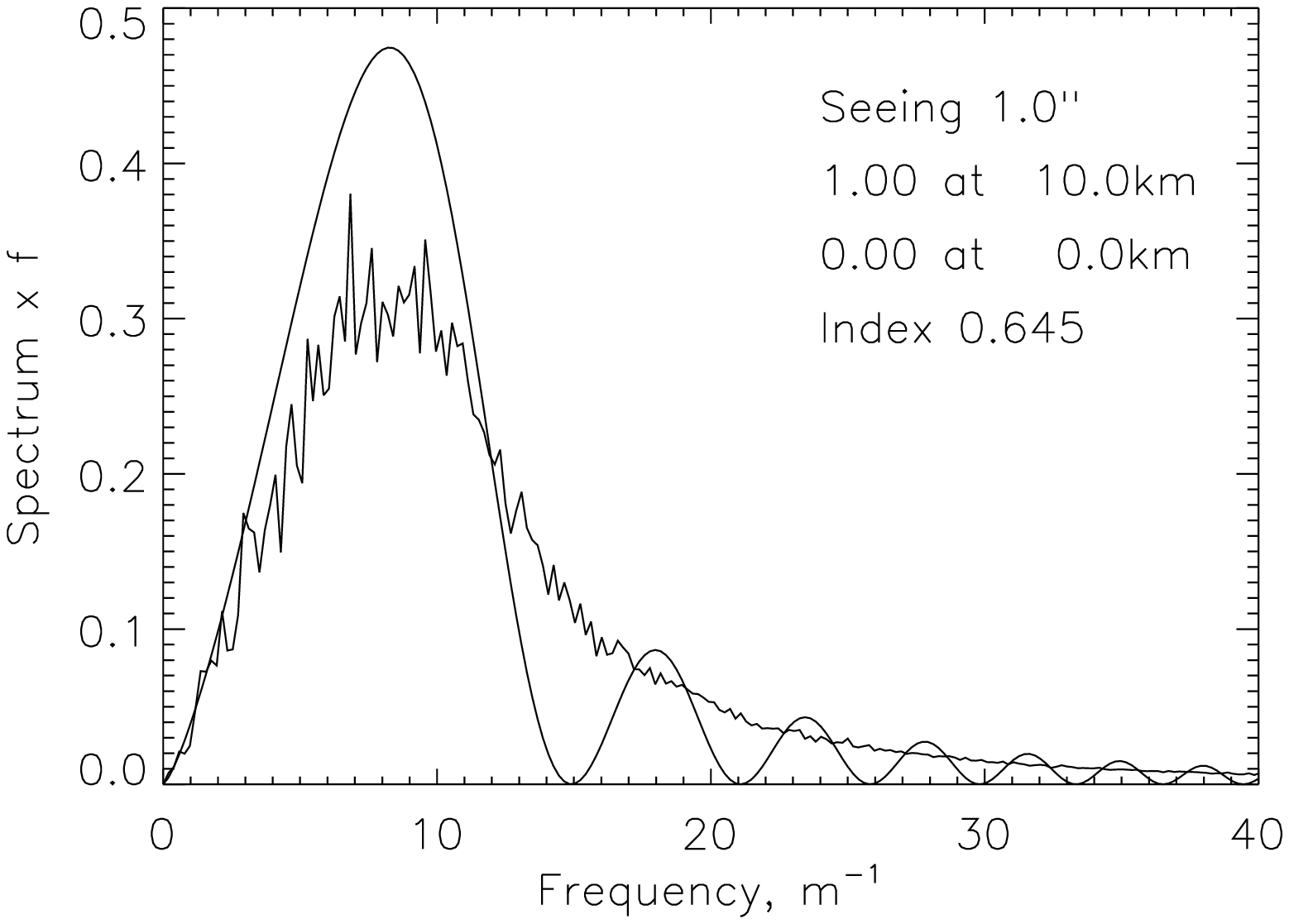} 
\caption{Radially-averaged    power     spectra    of    scintillation
  corresponding to the  two cases of Fig.~\ref{fig:intens}.  ``Noisy''
  curves  --  simulated, smooth  curves  -- weak-perturbation  theory,
  Eq.~\ref{eq:Wchi}.
\label{fig:power} }
\end{figure}

The equations (\ref{eq:Wphi}) and (\ref{eq:Wchi}) are only approximate
for a real (not weak) turbulence.  Even near the zenith, scintillation
indices $s^2 > 1$ were  measured.  This regime of strong scintillation
corresponds  to the  onset of  focusing when  the deviations  from the
standard  theory become  quite  significant.  Figures~\ref{fig:intens}
and  \ref{fig:power} show  simulated scintillation  signals  and their
spatial   power   spectra    for   weak   and   strong   scintillation
(cf.  Sect.~\ref{sec:simul}).   In the  latter  case,  the pattern  is
dominated  by  small  spikes due  to  focusing  of  the light  by  the
``lenses''  created by  the high-atmosphere  turbulence.  The  size of
these spikes is less than the size of the lenses and, accordingly, the
power  at   high  spatial   frequencies  increases  compared   to  the
weak-scintillation spectrum (\ref{eq:Wchi}),  while at low frequencies
it  decreases.   The  appearance  of  extra  power  can  be  explained
qualitatively in  the following  way.  Under weak  scintillation, each
sinusoidal component of the  phase screen creates intensity modulation
with the same  frequency.  As the amplitude of  the phase perturbation
increases, the phase gratings start to produce second and higher-order
or  crossed harmonics  in the  intensity distribution  at  the ground.
Individual layers  are no  longer independent but  cross-modulate each
other.  The  phase is affected by  the saturation in a  similar way as
the amplitude.

%--------------------------------------------------------------------
\subsection{Simulation tool}
\label{sec:simul}

Given the  lack of accurate  wave-propagation theory, the best  way to
study  higher-order propagation  effects is  by  numerical simulation.
Simulations are also  indispensable in evaluating various instrumental
effects.

We  use the  Fourier method  of generating  random phase  screens with
Kolmogorov  statistics. A  2-dimensional array  of  zero-mean Gaussian
complex random numbers is  created, their amplitudes $\propto \sqrt{C}
| {\bfit f}|^{-11/6}$  (${\bfit f}$ is  the spatial frequency,  $C$ is
the  coefficient from  Table~\ref{tab:seeing}) and  phases distributed
uniformly in  the interval $(-\pi, \pi)$.  The  Fourier transform (FT)
of this  array creates the phase  screen.  It is well  known that this
method underestimates the low-frequency components of turbulence, and,
notably, produces wrong phase  structure functions. The reason is that
any function obtained by discrete  FT is periodic, with a period equal
to the  grid size. However, the  effects of the  numerical outer scale
can  be neglected  if the  aperture size  is a  small fraction  of the
screen size.  Indeed, we  checked that in the weak-perturbation regime
our  numerical  simulations  reproduce  the  analytical  scintillation
spectra and DIMM response to within 2\%.

The  propagation   of  wave-fronts  is  calculated   by  the  spectral
technique. If $U_1({\bfit r})$ is the amplitude of the light wave before
propagation, and $\tilde{U}_1({\bfit f})$ is  its FT, then the FT of the
amplitude $\tilde{U}_2({\bfit f})$ after propagation over a distance $z$
is obtained by frequency filtering:

\begin{equation}
\tilde{U}_2({\bfit f}) = \tilde{U}_1({\bfit f}) \exp ( - i\pi z \lambda
|{\bfit f}|^2 ) .
\label{eq:Fresnel}
\end{equation}

This  method is  computationally fast,  involving only  two  FTs.  Its
drawback is that  in fact it describes the  propagation in a waveguide
with  a cross-section  equal  to the  simulation  grid and  reflective
walls.  However,  the propagation of  periodic phase screens  does not
present  problems  near  the  grid boundaries  because  the  amplitude
``wraps around''  the simulated domain boundaries and  does not produce
artifacts.

We generate  large screens of  complex light amplitudes at  the ground
resulting from the propagation through one or several turbulent layers
at  various altitudes.   Typically, with  1024$^2$ points  and 0.5\,cm
sampling the grid size is 5.12\,m.  To simulate the data sequence of a
seeing  monitor, the  screens are  shifted in  both coordinates  by $V
\tau$, where  $V$ is the  wind speed and  $\tau$ is the  sampling time
interval.  The shift  is directed at some small  angle with respect to
the $x$-axis.   In this way,  the aperture moves  mostly in $x$  but is
displaced in  $y$ by $\sim  0.2$\,m per line, eventually  covering the
whole screen one or several times. There are no adverse effects at the
borders because  the screens are periodic.   The statistical averaging
is sufficient to simulate typical 1-min measurements.

The complex amplitude of light at  the aperture of a seeing monitor is
re-sampled  on  a  finer  grid  and used  to  calculate  the  measured
quantities such  as the  spot images  in DIMM or  fluxes in  MASS. The
effects  of finite CCD  pixels and  detector noise  can be  studied as
well.
The current simulator  has some limitations. The wind  speed is common
to all layers.  The blur during a finite exposure time  in DIMM is not
simulated. In MASS,  we simulate the exposure time  $\tau$ by a linear
blur of  the apertures  over a distance  $V \tau$.  The  simulation is
usually  monochromatic.  

%We  simulate polychromatic  MASS data  by propagating  three different
%wavelengths through the same screens  and making a weighted sum of the
%intensities.

%------------------------------------------------------------------------
\section{Accuracy of the MASS method}
\label{sec:MASS}

%------------------------------------------------------------------------
\subsection{From scintillation to seeing}

The  MASS instrument  is based  on the  spatial analysis  of intensity
fluctuations  at   the  ground  level.   The  spatial   scale  of  the
scintillation ``speckle'' produced by  turbulence at a distance $z$ is
of  the order of  the Fresnel  radius $r_{\rm F}  = \sqrt{\lambda  z}$, i.e.
$\sim 10$\,cm  for a 10-km propagation  \citep{Roddier81}. The spectrum
(\ref{eq:Wchi}) reaches  maximum at the spatial frequency  $| {\bfit f}|
\sim r_{\rm F}^{-1}$.

Light  from a bright  star is  detected in  MASS with  four concentric
annular apertures with diameters  from 2\,cm (inner) to 8\,cm (outer).
The size  of these  apertures is of  the order  of $r_{\rm F}$ and  they act
jointly as a spatial  filter, permitting to dis-entangle scintillation
originating at different altitudes. This is achieved in several
steps.

{\bf Step  1.}  The sequences of  photon counts from  4 apertures with
individual  micro-exposure  of 1\,ms  are  acquired  and processed  to
calculate 10  {\em scintillation indices} --  normal indices $s^2_{\rm
A}$ for each aperture A  and 6 differential indices $s^2_{\rm AB}$ for
pairs of apertures A and  B.  The formulae for calculating the indices
and  subtracting the  bias caused  by the  photon noise  are  given in
\citep{Rest}.  The indices $s^2_{\rm  A}$ and $s^2_{\rm AB}$ are equal
to the variance  of the natural logarithms $\log  I_{\rm A}$ and $\log
(I_{\rm A}/I_{\rm B})$ in the limit of small fluctuations of the light
intensities $I_{\rm A}$ and $I_{\rm  B}$, $s^2 \ll 1$.  The turbulence
theory usually operates with  the variance of the logarithm.  However,
we measure the  photon counts which can be zero  and fluctuate even at
constant light, hence the indices should be calculated from the normal
(non-logarithmic) variances.

{\bf Step 2.} A linear relation  between the observables (indices) and the
   turbulence  profile $C_n^2 (z)$   is  established by  the
   weak-perturbation theory,
\begin{equation}
s^2_k = \int W_k(z)\; C_n^2 (z) {\rm d} z ,
\label{eq:WF}
\end{equation}
where  the  {\it  weighting  function}  (WF)  $W_k(z)$  describes  the
altitude  response of a  given aperture  or aperture  combination $k$.
For a  weak Kolmogorov  turbulence, the WF  depends {\em only}  on the
aperture  geometry  and spectrum  of  detected  radiation  and can  be
derived from (\ref{eq:Wchi}) \citep{AO02,Poly}.   A normal index for an
aperture of  diameter $D_{\rm A}$ corresponds  to the low-pass  filtering of
scintillation,  passing   $|  {\bfit  f}  |  <   1/D_{\rm A}$,  whereas  a
differential  index corresponds  to  a band-pass  spatial filter  with
$1/D_{\rm B} < |  { \bfit f} | < 1/D_{\rm A}$.  Thus,  MASS with its centimetric
apertures is  sensitive to the turbulence of  centimetric scales where
the Kolmogorov model is adequate.

It  has been  shown  that  the differential  index  in two  concentric
annular apertures  is almost  independent of the  propagation distance
$z$ for $z> z_{\rm AB} = D_{\rm AB}^2/\lambda$, where $D_{\rm AB} = (D_{\rm A} + D_{\rm B})/2$
is the average aperture  diameter \citep{AO02,Poly}.  It means that the
scintillation index gives a direct measure of the turbulence integral,
hence  seeing, produced at  distances beyond  $z_{\rm AB}$.  On  the other
hand, normal scintillation indices  increase as $z^\beta$, with $\beta
= 5/6$ for small distance $z \ll D^2/\lambda$ and $\beta=2$ for $z \gg
D^2/\lambda$ \citep{Roddier81}. Accordingly,  most of the scintillation
is produced by the high layers.

{\bf Step  3.} Using the  known WFs, the  set of 10 indices  is fitted
with a model of 6 thin  turbulent layers at altitudes $h_i$ of 0.5, 1,
2, 4, 8, and 16\,km, with  turbulence integrals $J_i$ in each layer as
parameters  \citep{Rest}.  The  zenith  angle $\gamma$  is taken  into
account by the model, $z = h \sec \gamma$.  In reality, the turbulence
is distributed  continuously in altitude  with a profile  $C_n^2 (h)$.
The integrals  $J_i$ delivered  by MASS are  approximately equal  to $
\int C_n^2  (h) R_i(h) {\rm  d}h$, where the {\it  response functions}
$R_i(h)$ resemble  triangles in $\log h$ coordinate  centred on $h_i$
\citep{Rest}.   The sum  of all  $R_i(h)$  is close  to one  for $h  >
0.5$\,km.

%------------------------------------------------------------------------
\subsection{Over-shoots and their correction}

\begin{figure*}
\centerline{\includegraphics[width=8cm]{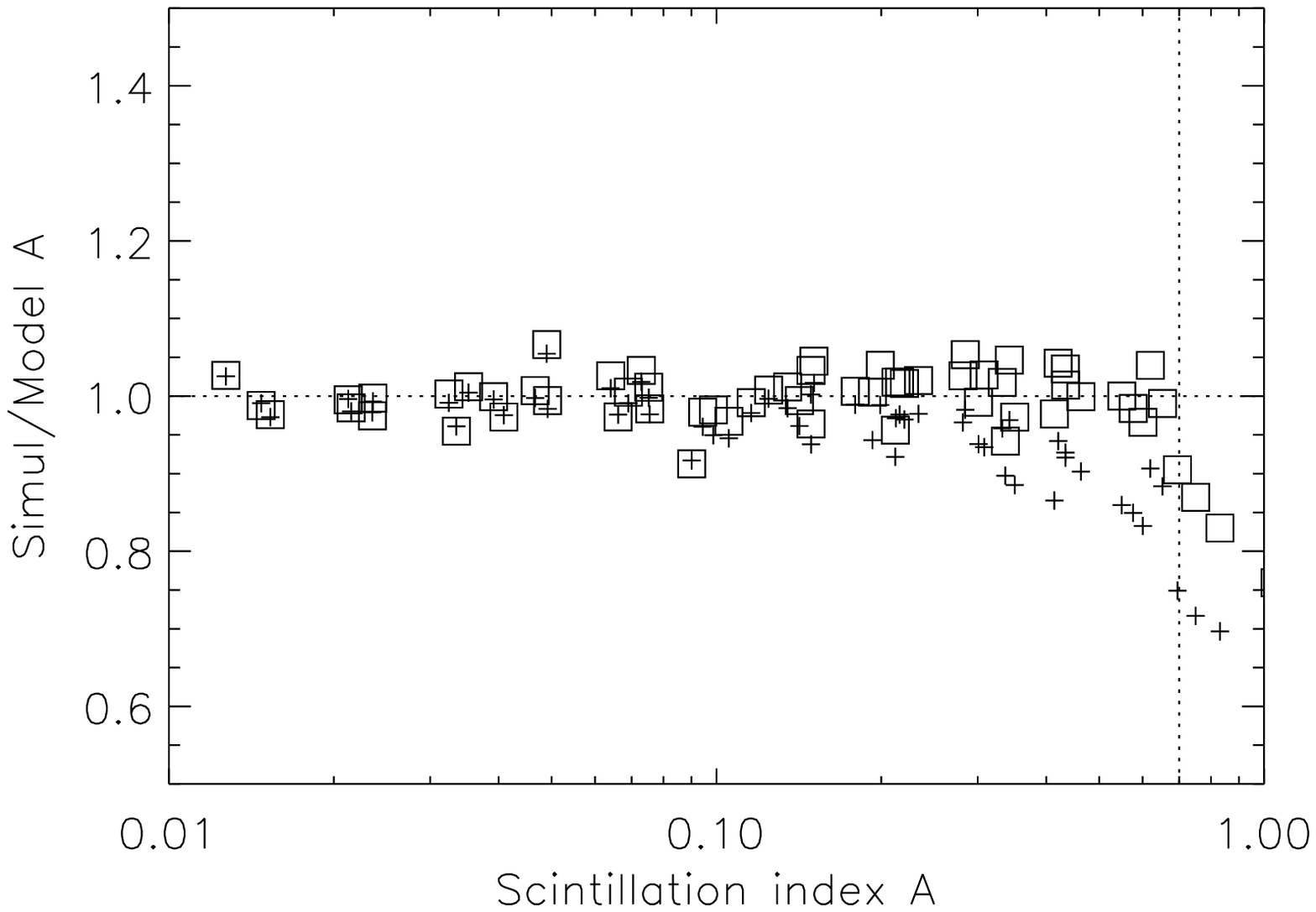} 
\includegraphics[width=8cm]{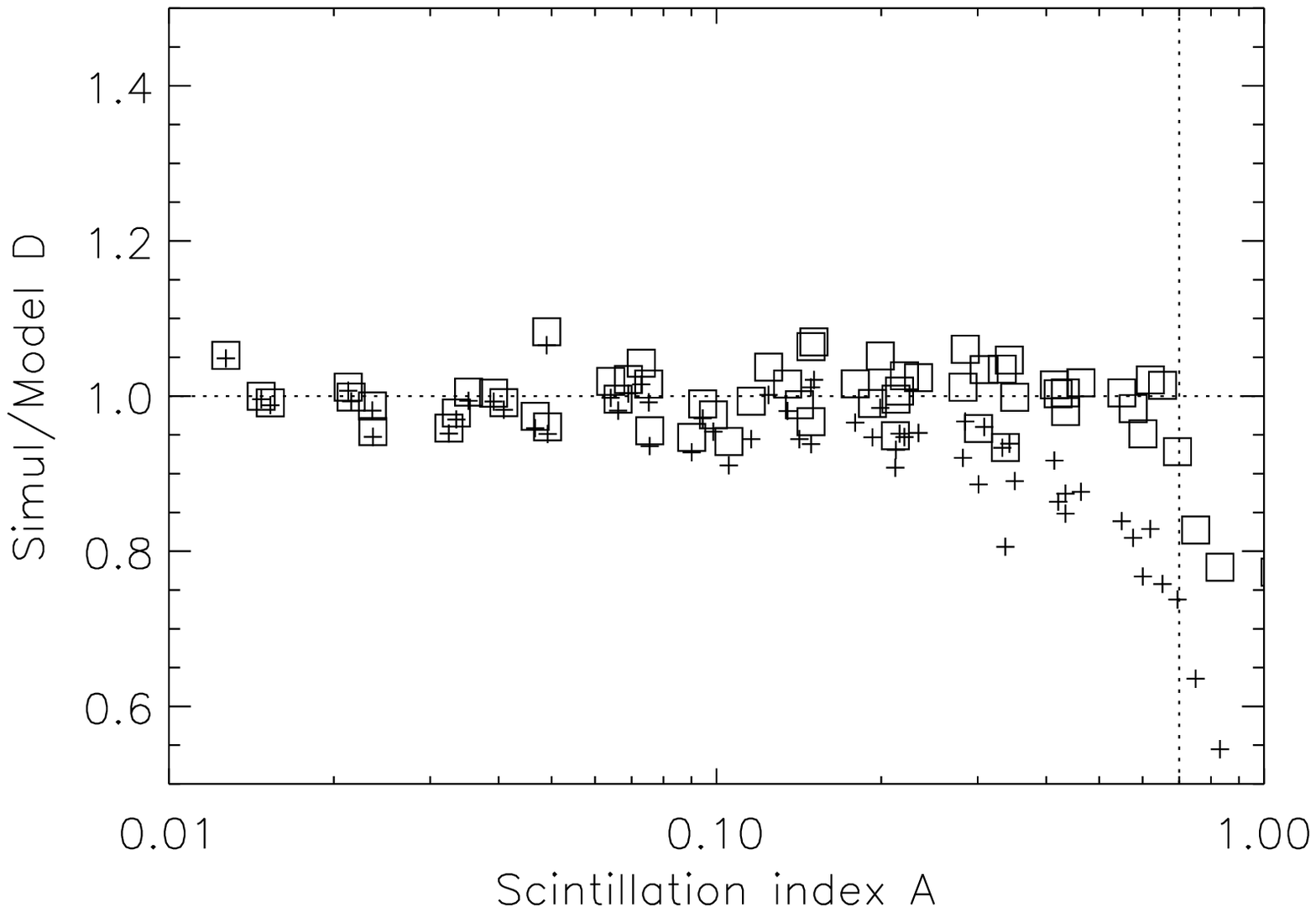}   }
\centerline{\includegraphics[width=8cm]{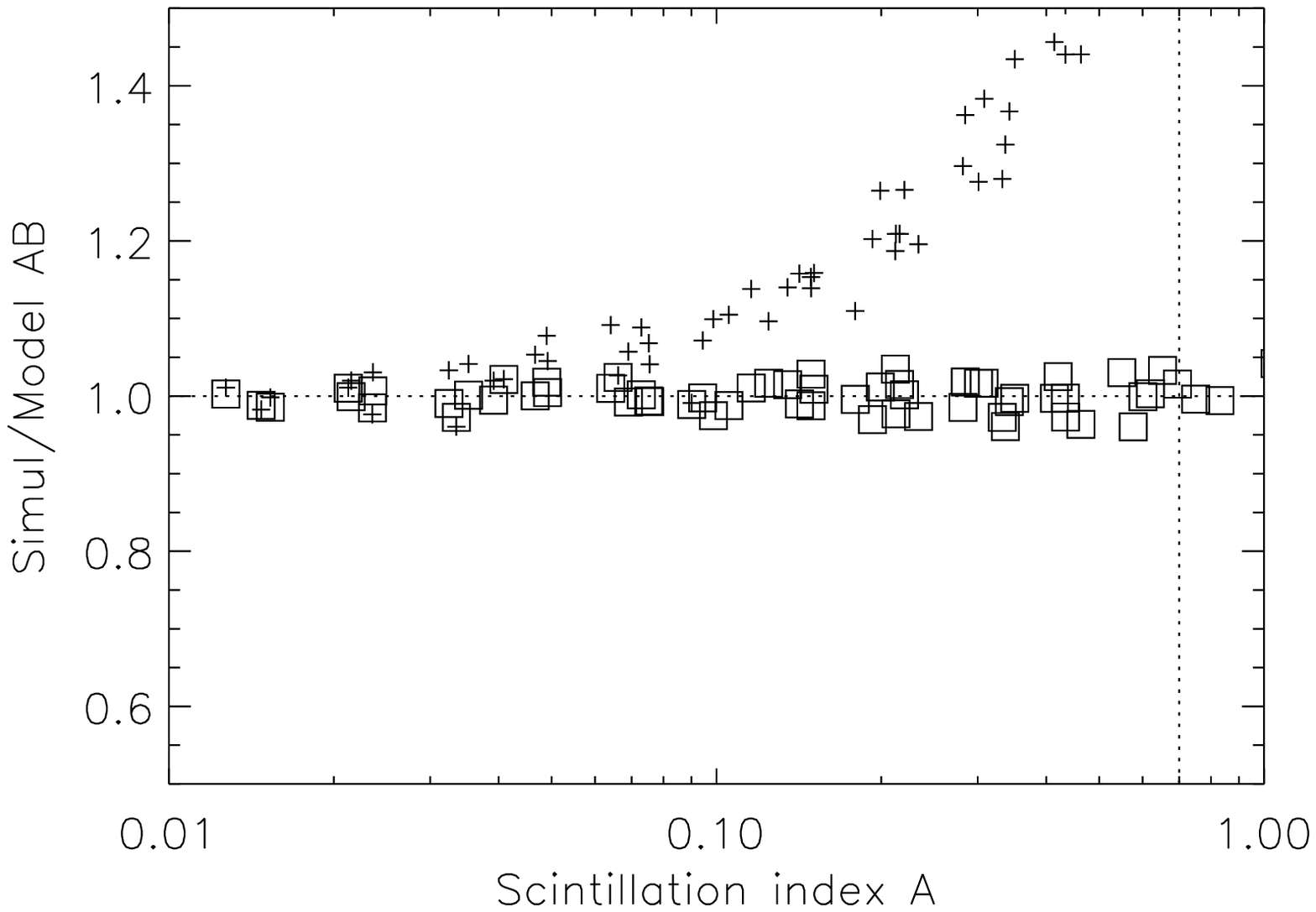} 
\includegraphics[width=8cm]{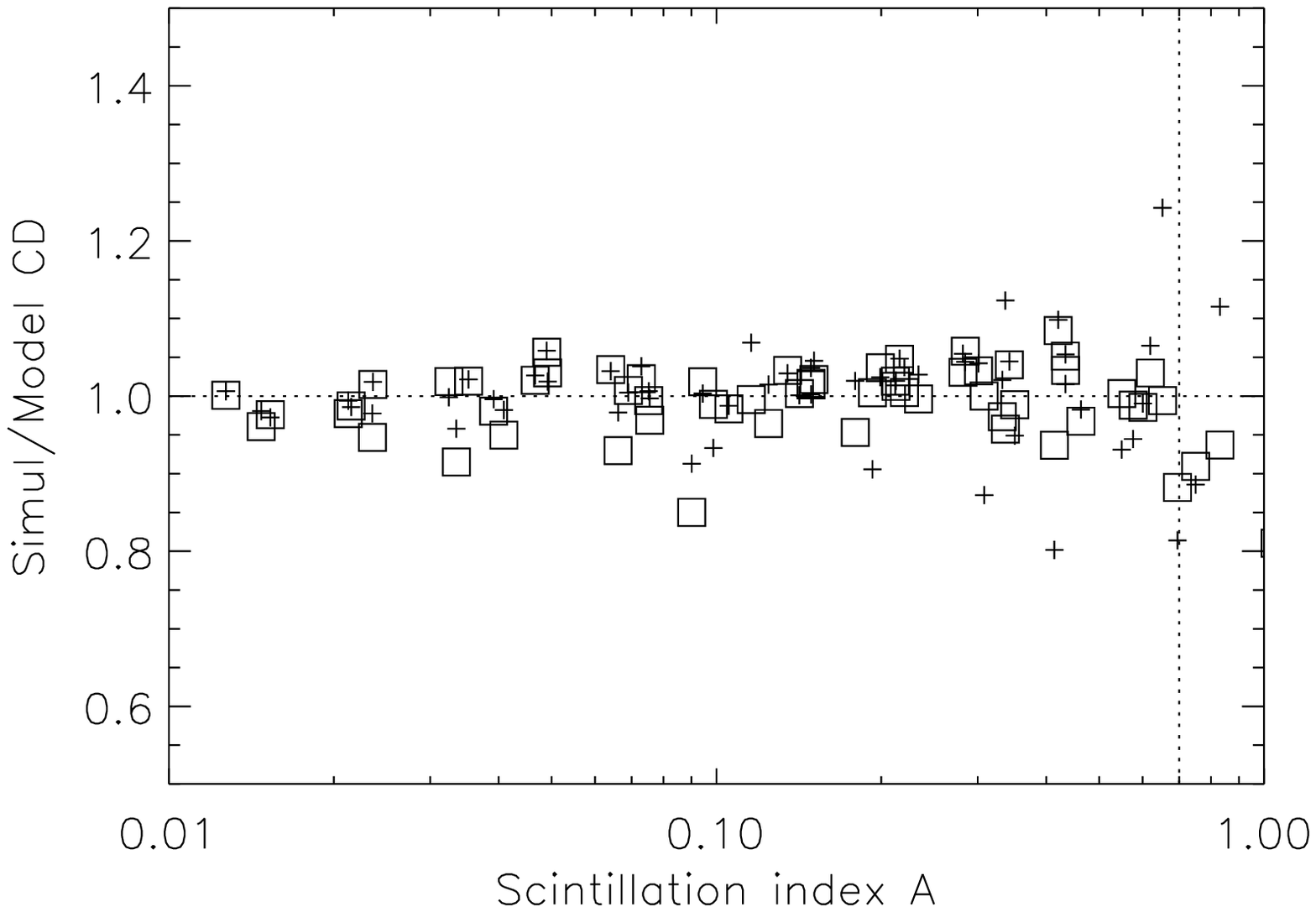}   }
\caption{Ratio of measured to modelled scintillation indices.  Selected
 normal (top row)  and differential (bottom row) indices  in some MASS
 apertures as determined from the simulations without correction ($s^2$,
 crosses) or after correcting by Eq.~\ref{eq:S} ($s_*^2$, empty squares)
 are  divided by  the  weak-scintillation indices  $s_0^2$.  The  dotted
 vertical line marks the correction limit $s_{\rm A}^2=0.7$.  }
\label{fig:modlin}
\end{figure*}

MASS  relies  on  the  small-perturbation theory,  assuming  that  the
scintillation is  weak, $s^2 \ll 1$,  and that the  combined effect of
several turbulent layers  is additive (Eq.~\ref{eq:WF}).  In practise,
scintillation  is  not always  weak  (Sect.~\ref{sec:prop}).  In  this
case,  profile restoration  by the  linear method,  as  implemented in
MASS,   leads  to   the  over-estimated   total   turbulence  integral
(free-atmosphere seeing) and  to the shift of the  restored profile to
lower altitudes.

Although the theory of strong scintillation has been addressed in many
papers    \citep[e.g.][and references therein]{A99},  there  is  no
quantitative  description of the  intensity power  spectrum available.
In order to extend  MASS operation to moderately strong scintillation,
we      rely       exclusively      on      numerical      simulations
(Sect.~\ref{sec:simul}).  Poly-chromatic light  was simulated by an
equal mix of  three wavelengths of 0.4, 0.45, and  0.55 $\mu$m. One or
two phase  screens at various altitudes were  simulated.  The physical
size of simulated phase screens is 2.5\,m across, with 2.5\,mm pixels.
The  power  spectrum  of  intensity  fluctuations at  the  ground  was
computed  (e.g. Fig.~\ref{fig:power}) and  converted into  normal and
differential  scintillation indices  of  MASS by  integrating it  with
suitable spatial filters.

The  values of  the measured   indices  inform us  on the
strength of the scintillation. Our approach is to correct the measured
indices  semi-empirically and to  bring them  into agreement  with the
weak-scintillation theory.  Let ${\bfit s^2}$ be the 10-element vector
of  measured indices,  ${\bfit s}^2_0$  -- the  vector  of theoretical
indices expected in the  linear theory without saturation, and ${\bfit
s}^2_*$  --  the  vector  of  corrected  indices.   A  rather  general
correction formula can be written as

\begin{equation}
{\bfit s}^2_* = \frac{{\bfit s}^2} { 1 + {\bfsf Z} {\bfit s}^2} \approx {\bfit s}_0^2 .
\label{eq:S}
\end{equation}
The  rationale for  selecting this  formula is  that  it automatically
removes  correction for  weak  scintillation, and  that  this type  of
formula works well for the  differential indices. Here, {\bfsf Z} is the $10
\times 10$ {\em correction matrix}.

We determine the correction matrix {{\bfsf Z} empirically from the results of
simulations, by  least-squares fitting.  The fitting  is restricted to
the  relevant scintillation  range $  0.1 <  s^2_{\rm A} <  0.7$  because weak
scintillation does  not need correction (but  adds statistical noise),
while correcting  stronger scintillation is hopeless. The  matrix {\bfsf Z} is
found by least squares as 
\begin{equation}
{\bfsf Z} = ( {\bfsf S} {\bfsf S}^T)^{-1} ( {\bfsf S}^T {\bfsf Y}), \;\;\; 
{\bfsf Y} = {\bfsf S}/{\bfsf S}_0 -1 ,
\label{eq:Zmat}
\end{equation}
where  the  matrices  {\bfsf  S},  ${\bfsf  S}_0$,  ${\bfsf  Y}$  have
dimensions $10 \times M$, $M$  being the number of simulated cases. We
simulated  both single  and double  layers at  various  altitudes with
various strengths of  each layer.  All results were  mixed together in
calculating  the  matrix {\bfsf  Z},  with  a  total of  $M=35$  cases
satisfying the condition $ 0.1 < s^2_{\rm A} < 0.7$.
 
Figure~\ref{fig:modlin} demonstrates the success of this approach with
plots  of  the  simulated  (pluses) and  corrected  (squares)  indices
divided   by   the  weak-scintillation   ones.    The  average   ratio
$s^2_*/s^2_0$ of  all indices  differs from 1  by less than  1\% after
correction,  the  scatter is  also  significantly  reduced.  The  best
correction  is  achieved  for  $s^2_{\rm AB}$, rms  $s^2_*/s^2_0$  scatter
1.6\%, and the worst correction -- for $s^2_{\rm D}$, rms scatter 4.6\%.

\begin{figure}
\includegraphics[angle=270,width=8cm,]{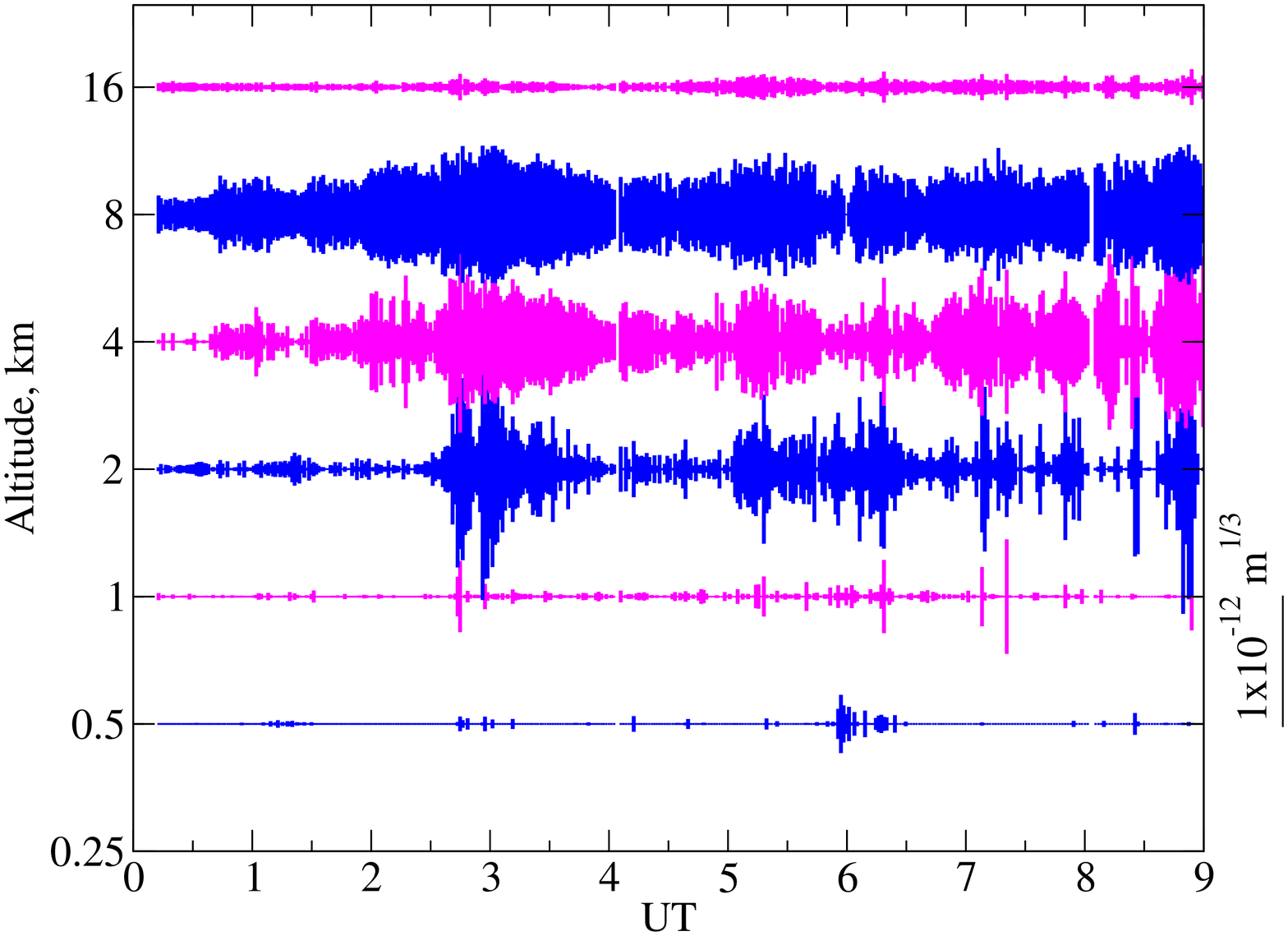} 
\includegraphics[angle=270,width=8cm]{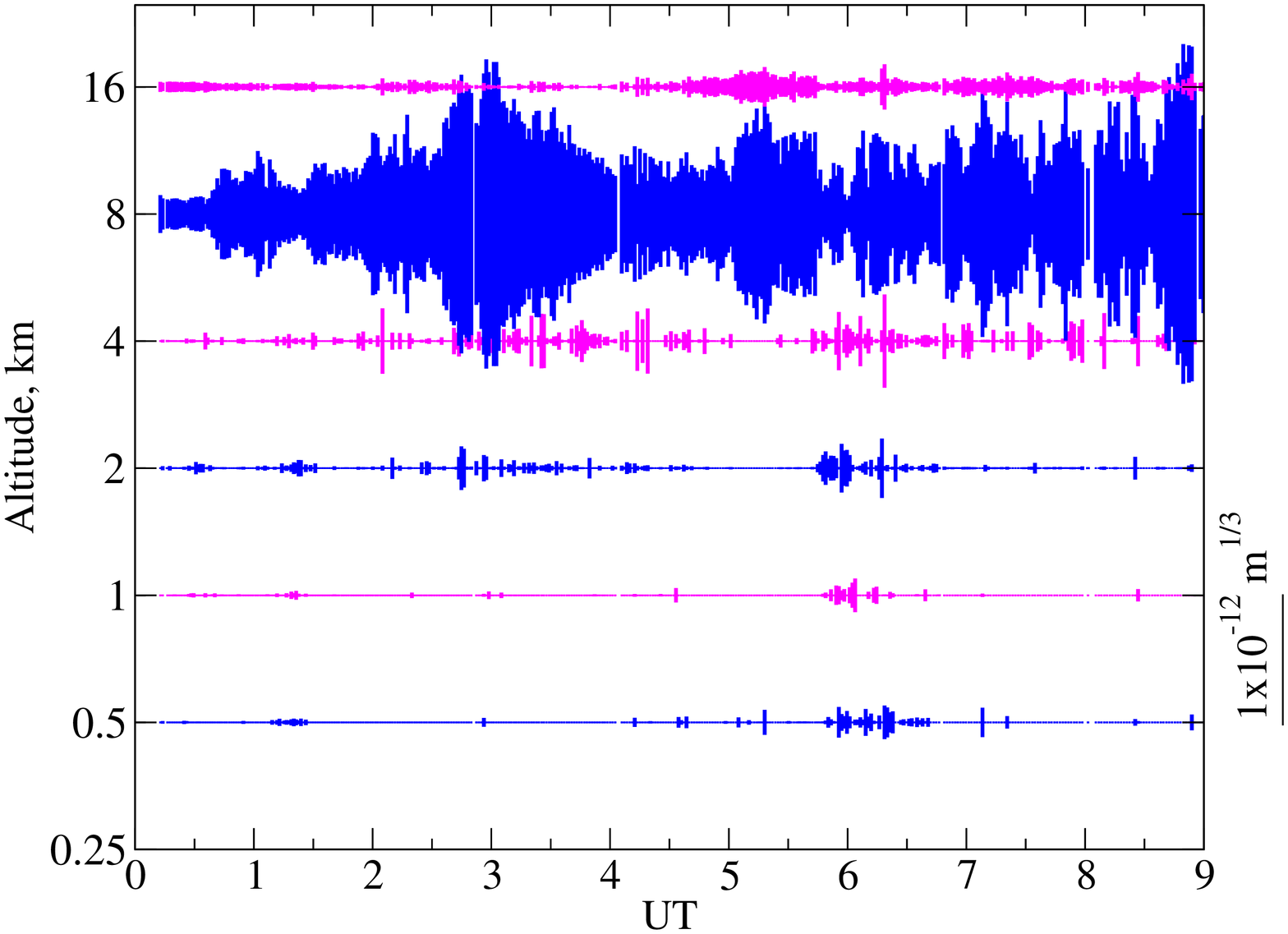} 
\includegraphics[angle=270,width=8.5cm]{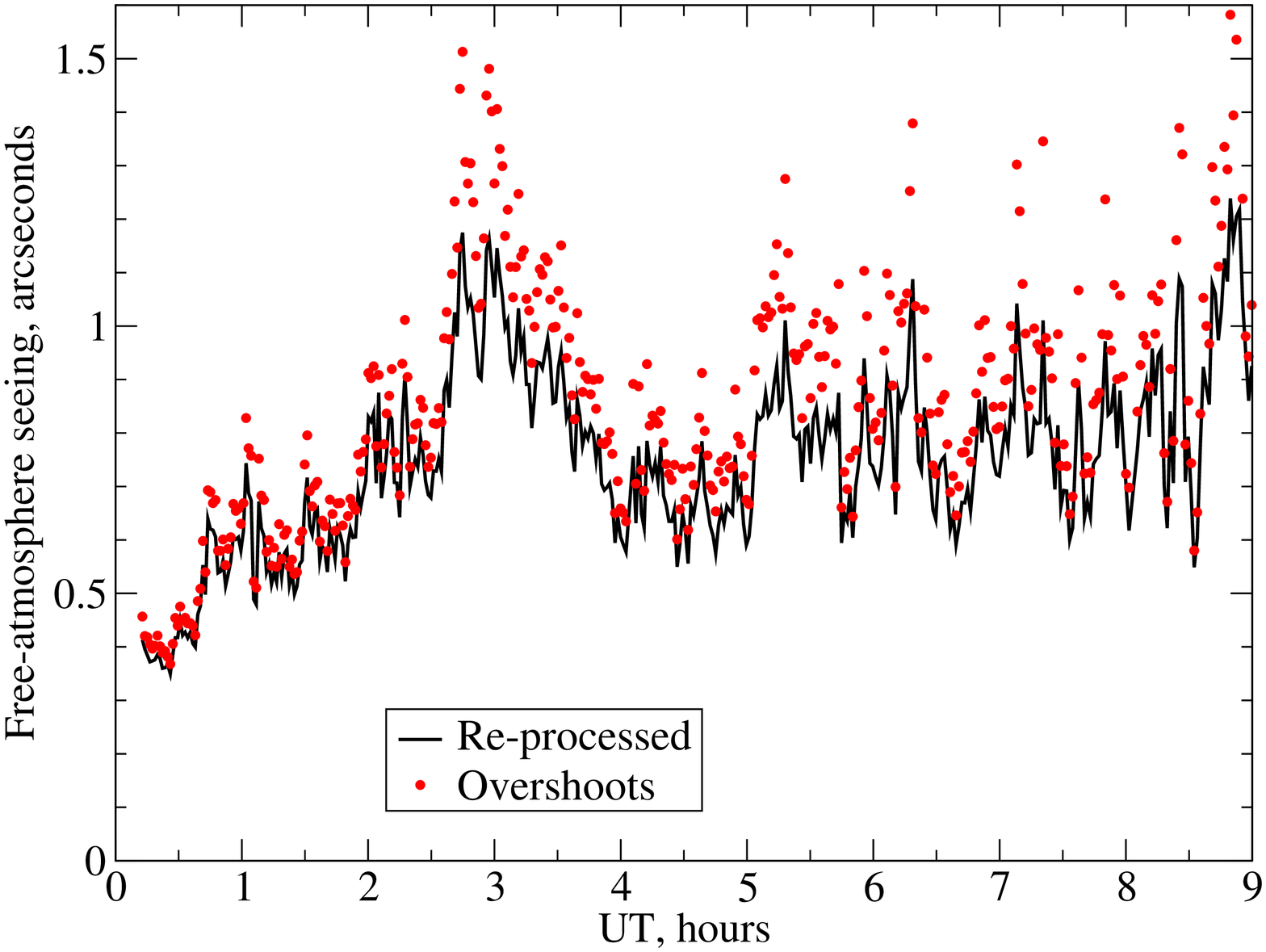}
\caption{Example  of MASS data  re-processing. Turbulence  profiles at
Cerro Tololo for  the night of September 30, 2004  are plotted as bars
with length proportional to the integrals $J_i$ (cf. the scale on the
right).  Top -- original results, middle -- with over-shoot correction.
The free-atmosphere seeing calculated  from these profiles is compared
on the lower plot.
\label{fig:reproc} }
\end{figure}

The  correction  of  indices  with  the matrix {\bfsf Z} ``learned''  from  the
simulations  is  implemented  in  the  current  version  of  the  MASS
software, {\small TURBINA}.   Old data can  be re-processed with  this program.
An  example  of  successive   ``overshoot''  correction  is  shown  in
Fig.~\ref{fig:reproc}.  Our  empirical correction technique  works for
$s^2_{\rm A}< 0.7$. The  correction matrix is determined only  for the typical
aperture diameters and bandpass used in  MASS, it has to be revised if
these parameters change.

A simplified method of correcting MASS seeing for over-shoots has been
established  earlier and  works quite  well. If  $r_{0, MASS}$  is the
Fried parameter of the free-atmosphere seeing measured with the old MASS
software, then the corrected $r_0$ will be
\begin{equation}
r_0 \approx r_{0, MASS} \;(1 + 0.7 s_{\rm A})^{0.6} .
\label{eq:AB}
\end{equation}

%------------------------------------------------------------------------
\subsection{Temporal sampling} 

The photon counts in MASS are sampled with the exposure time $\Delta t
=1$\,ms.  Averaging   the  signals  during  $\Delta   t$  reduces  the
fluctuations, causing  a bias  in the measured  indices. This  bias is
corrected  in the software  by calculating  the indices  with exposure
time $\Delta  t$ and $2 \Delta  t$ and extrapolating  linearly to zero
exposure.  This  extrapolation is calculated  as $s_0^2 = 1.5  s_1^2 -
0.5 \rho_1$, where  $s_1^2$ is the measured index  and $\rho_1$ is the
covariance with a time lag of 1 sampling period \citep{Rest}.

In  October 2004  we  recorded  MASS signals  with  a faster  sampling
$\Delta t =0.25$\,ms under  rapid-turbulence conditions.  It turned out
that  the  linear  extrapolation  actually over-corrects  the  indices
$s_{\rm A}^2$ and $s_{\rm AB}^2$ by as  much as $\sim 6$\%.  An analytical study
made earlier \citep{AO02} also concluded that a less drastic correction
works better.  Accordingly, in the new MASS software the correction is
halved, $s_0^2 =  1.25 s_1^2 - 0.25 \rho_1$.  The  remaining bias is under
2\% even on the fastest indices $s_{\rm A}^2$ and $s_{\rm AB}^2$.

%------------------------------------------------------------------------
\subsection{Inner turbulence scale}

\begin{figure}
\centerline{\includegraphics[width=8cm]{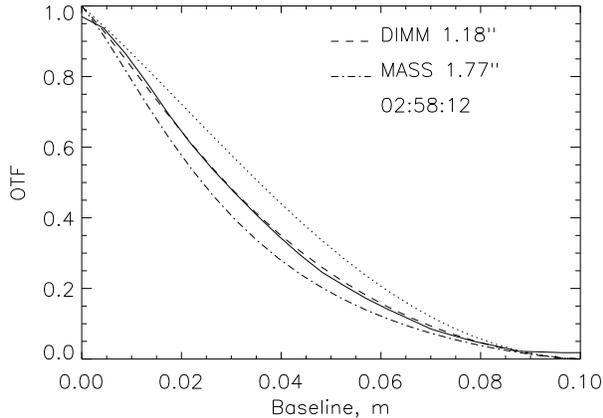} }
\caption{Comparison  of the  optical transfer  function of  DIMM spots
  (full line) with the models based on the Kolmogorov spectrum and the
  seeing  measured by  the DIMM  (dashed line)  and  MASS (dash-dotted
  line)  channels. Cerro Tololo,  October 1,  2004 2:58UT.  The dotted
  line is a diffraction-limited OTF.
\label{fig:otf} } 
\end{figure}

Turbulence  spectrum at  high spatial  frequencies (comparable  to the
turbulence  inner  scale)  may  have  excess  power  compared  to  the
Kolmogorov model,  the so-called {\em Hill bump}  \citep{A99}.  The inner
scale is  usually of the  order of few  millimetres, but it  can reach
centimetric  values  in the  upper  atmosphere.   This phenomenon  can
potentially increase the scintillation and lead to over-shoots.

Apart from the extra  scintillation, small distortions unaccounted for
by the Kolmogorov model must  cause excessive image blur, appearing as
an extra halo of the PSF.  We tried to detect such effect by analysing
average  re-centred  images  of a  star  in  the  DIMM channel  of  a
MASS-DIMM.   With   10-cm  DIMM   apertures,  the  spots   are  always
diffraction-limited  under good  seeing, but  become broadened  as the
seeing  degrades.   Figure~\ref{fig:otf}   shows  an  example  of  the
1-dimensional optical  transfer function (OTF)  of average re-centred
DIMM spots in spatial coordinate $r  = \lambda f$.  It is compared to
the product of  the short-exposure atmospheric and diffraction-limited
TFs  calculated for $\lambda  = 0.55$\,$\mu$m  with $r_0$  measured by
both MASS  and DIMM.   We do  not notice any  departure from  the DIMM
model  at centimetric  scales and  conclude that  the  $1.18''$ seeing
adequately described the spot profile, without any perceptible effects
of  the ``Hill  bump''.  On  the other  hand, MASS  over-estimated the
seeing  because of  the un-corrected  over-shoots.  This  analysis was
repeated several times  on other nights.  We conclude  that the effect
of the finite inner scale on MASS can be safely neglected.

%-----------DIMM-------------------------------------------------------------
\section{Accuracy of the DIMM method}
\label{sec:DIMM}

DIMM  has a  reputation of  a ``fool-proof''  technique  for measuring
seeing.  However, this is not true because its results are affected by
optical propagation,  centroid algorithm, aberrations,  exposure time,
noise, etc.  Some  of these effects can cause a  bias much larger than
10\%.   The accuracy  of  the DIMM  method  is studied  below by  both
numerical simulations  and analytical theory. The  analytics is useful
for understanding the small-signal response  of a DIMM which turns out
to be  rather different from  the usual assumption that  DIMM measures
tilts.  On  the other  hand,  simulations  reveal  the faults  of  the
weak-perturbation theory, as for the MASS. 

We consider  here a DIMM instrument with  typical parameters: aperture
diameter  $D=10$\,cm  and   baseline  $B=25$\,cm.   In  the  numerical
simulations, we selected  a pixel scale of 0\farcs32  and a field size
of $10''$ around each spot. The complex amplitude of the monochromatic
($\lambda =  0.5\,\mu$m) light waves  at the ground  after propagation
through    turbulence    was    calculated   with    1\,cm    sampling
(Sect.~\ref{sec:simul}).  The same  amplitude screens  can  be re-used
with  varying  DIMM  parameters  (e.g.  aberrations).   The  simulated
baseline  is 0.24\,m  (even number  of pixels).   The  centroid window
follows the  spot, as in  typical DIMM instruments,  because otherwise
the spots would move with respect to the window and the response under
bad seeing  would be  diminished.  With 4000  spot pairs in  a typical
simulation, the statistical error  of the differential variance (hence
of  the  derived response  coefficients)  is  1.6\%.   Apart from  the
statistical  errors,   subtle  details   of  the  algorithm   and  its
implementation may influence the results of simulations, which are not
{\em exact} in the absolute sense.

%------------------------------------------------------------------------
\subsection{Response coefficient of an ideal DIMM}
\label{sec:resp-ideal}

 In a DIMM, two circular  portions of the wavefront are isolated.  The
variance  of   the  differential  wave-front   tilts  in  longitudinal
(parallel  to the base)  $\sigma^2_{l}$ and  transverse $\sigma^2_{t}$
directions   is   related   to    the   Fried   parameter   $r_0$   as
\citep{Martin87,DIMM,PASP02}
\begin{equation}
\sigma^2_{l,t} = K_{l,t} \; (\lambda/D)^2 \; (D/r_0)^{5/3} .
\label{eq:KDIMM}
\end{equation}

The {\it  response coefficients} of  DIMM $K_{l}$ and $K_t$  depend on
the $B/D$  ratio and on the  kind of the tilt  measured.  Usually, the
tilt is evaluated  from the centroids of two  stellar images formed by
the  sub-apertures; in  this  case it  corresponds  to the  wave-front
gradient  (G-tilt). The  response  coefficients for  the  G-tilt as  a
function  of   $b  =  B/D$   can  be  approximated  by   the  formulae
\citep{PASP02}
\begin{equation}
 \begin{array}{l}
K_{\rm l} = 0.340 \; (1 - 0.570 b^{-1/3} -0.040  b^{-7/3})   \\
K_{\rm t} = 0.340 \; (1 - 0.855 b^{-1/3} +0.030  b^{-7/3}) . \\ 
\end{array}
\label{eq:K}
\end{equation}
For our example, $b=2.5$, $K_l = 0.1956$, and $K_t = 0.1270$.

The  $r_0$  parameter  (and  seeing)  is computed  from  the  measured
differential   image-motion  variance   $\sigma^2_{l}$   by  inverting
Eq.~\ref{eq:KDIMM}.  Hence,   it  is  proportional   to  the  response
coefficients to the power $3/5$. A 10\% error in the response entails
a 6\% error in seeing. 

%------------------------------------------------------------------------
\subsection{Response of  the centroid algorithm}
\label{sec:resp-real}

\begin{figure}
\includegraphics[width=8cm]{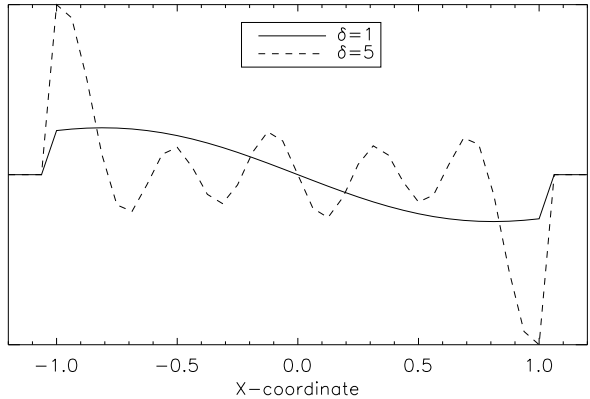} 
\caption{Cuts through the  centroid filter functions $F_\varphi({\bfit
  x})$  in the $x$-direction  are plotted  in arbitrary  units against
  normalised $x$-coordinate,  where $\pm  1$ corresponds to  the pupil
  border.    Two  centroid   windows:  narrow   $\delta=1$   and  wide
  $\delta=5$.  No aberrations, $D=0.1$\,m.
\label{fig:mask}}
\end{figure}

In  a  real  DIMM  instrument,   the  image  motion  is  estimated  by
calculating centroids of the spots.  Only a sub-set of detector pixels
is used  in order to reduce  the influence of the  noise. These pixels
are selected either  by setting a threshold well  above the background
noise  or by  defining  a  window around  the  brightest pixel.   Both
approaches can be expressed by a general formula
\begin{equation}
c_x = \sum_{i,j} w_{i,j} x_{i,j} I_{i,j} 
/ I_0, \;\;\;  
I_0 = \sum_{i,j} w_{i,j} I_{i,j},   
\label{eq:centr}
\end{equation}
where $c_x$  is the  estimated centroid $x$-coordinate,  $I_{i,j}$ are
pixel intensities,  $x_{i,j}$ are their  $x$-coordinates.  The weights
$w_{i,j}$ equal one for selected pixels and zero otherwise, although a
more sophisticated weighting scheme could be adopted.  Here we explore
circular  windows where  $w=1$  for  pixels at  a  distance less  than
$\delta  \lambda/D$  from the  spot  centre.   The parameter  $\delta$
defines the radius of the  centroid window in units of the diffraction
spot size $\lambda/D$.

Formula (\ref{eq:centr})  is only an  approximation to the  true image
centroid.   As a result,  the response  coefficient of  a DIMM  is not
exactly equal to its theoretical  value for G-tilt, but rather depends
on the  details of centroid  calculation.  We developed  an analytical
formula relating the distribution of  the light phase and amplitude at
the  pupil  to   the  change  of  the  image   centroid  $\Delta  c_x$
(Appendix~\ref{sec:PResp}),  valid  for  small perturbations.   It  is
analogous to  the Taylor expansion of PSF  developed for high-contrast
imaging \citep{Perrin}. The centroid shift $\Delta c_x$ is equal to the
sum of  the integrals of atmospheric  perturbations $\varphi({\bfit x})$
and  $\chi ({\bfit  x})$ multiplied  by  the {\em  filter functions}  $
F_{\varphi}    ({\bfit   x})$    and   $    F_{\chi}   ({\bfit    x})   $,
respectively. Knowing  the spectra of the  perturbations, we calculate
the small-signal response coefficients of a DIMM with Eq.~\ref{eq:Kl}.
The  relative  accuracy of  our  calculation is  3\%  or  less. 

Figure~\ref{fig:mask}  shows  the  filter  functions of  the  centroid
estimator   (\ref{eq:centr})  in   the  pupil   plane   calculated  by
Eqs.~\ref{eq:F},\ref{eq:A}.    For  a  narrow   ($\delta=1$)  centroid
window, the filter resembles remotely a Zernike tilt, while for a wide
($\delta=5$)  window it  is closer  to  a gradient  averaged over  the
aperture (opposite spikes  at the edges).  The DIMM  optics is assumed
perfect,   therefore  the   centroid  is   insensitive   to  amplitude
fluctuations at the pupil (scintillation), $ F_{\chi} ({\bfit x}) =0$.

\begin{figure}
\includegraphics[width=8cm]{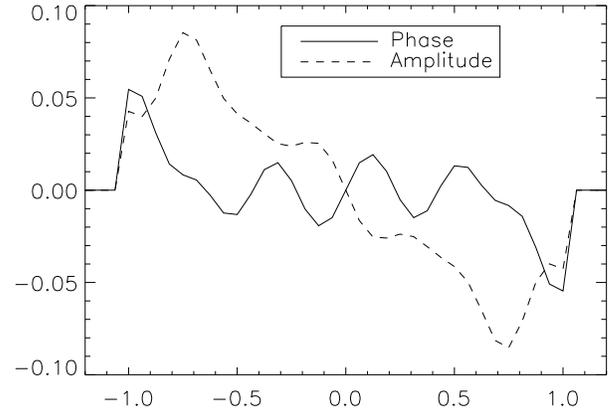}  
\caption{Centroid   filter   functions   $F_\varphi({\bfit  x})$   and
  $F_\chi({\bfit x})$  for a defocus of 1\,radian  rms and $\delta=5$.
  Compare with Fig.~\ref{fig:mask}.
\label{fig:def1}}
\end{figure}

\begin{figure}
\includegraphics[width=8.5cm]{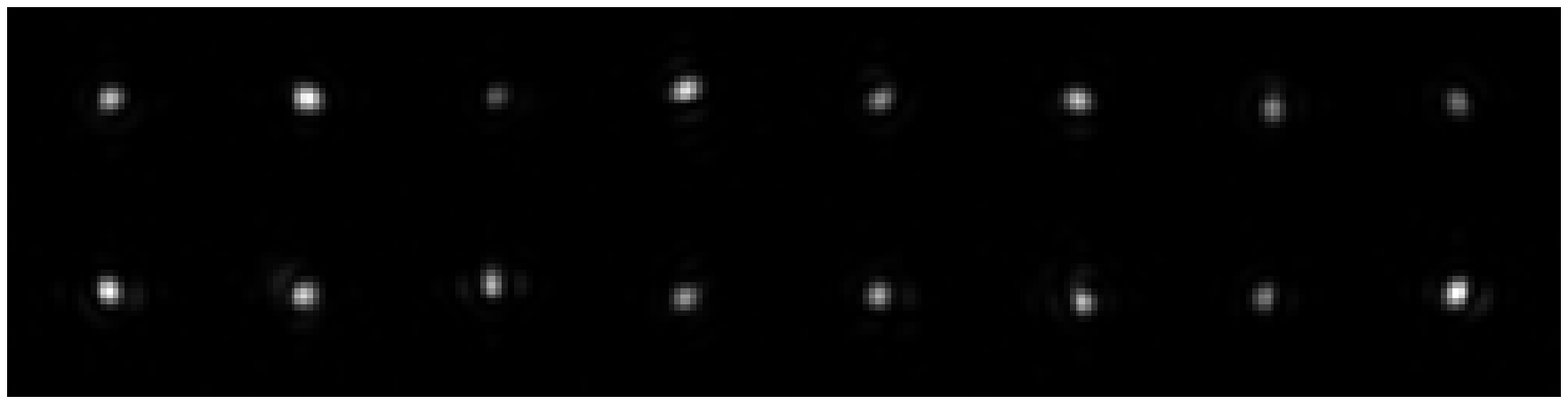} 
\medskip
\includegraphics[width=8.5cm]{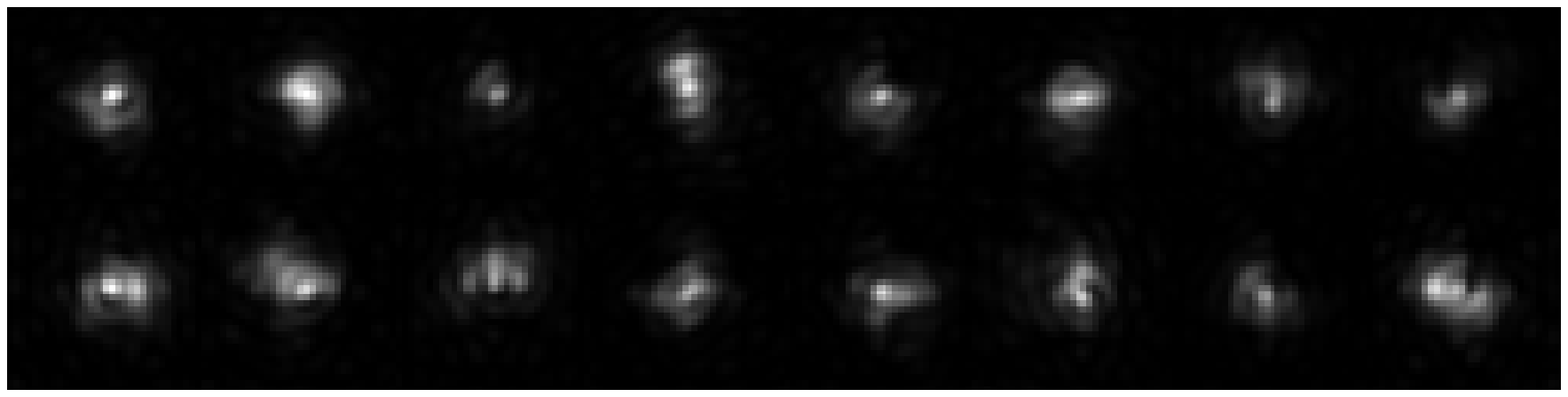}  
\caption{Sequences  of simulated  spot pairs  produced by  a turbulent
  layer at  10\,km with $r_0 =  0.1$\,m in a DIMM  with perfect optics
  (top) and with a defocus  of 1\,rad (bottom).  The same phase screen
  is used in both cases.
\label{fig:mosaic}}
\end{figure}

The situation  becomes more complicated  when we consider  a realistic
DIMM  instrument with  some optical  aberrations.  In  this  case, the
centroid $c_x$ is affected by both phase and amplitude fluctuations at
the  pupil  (Fig.~\ref{fig:def1}).   This  effect  can  be  understood
qualitatively:  a defocused spot  resembles a  pupil image,  hence the
centroid  estimator  becomes  sensitive   to  the  gradient  of  pupil
illumination.      Sequences      of     simulated     spot     images
(Fig.~\ref{fig:mosaic})  illustrate this  situation.  In  a  DIMM with
perfect optics, the spots are sharp, while their intensities fluctuate
because of the scintillation.  The defocused spots look more distorted
by  the  same  seeing  and  their centroids  move  more,  biasing  the
measurements.

The aberrations are characterised here by the amplitude of the Zernike
polynomials representing  the phase  {\em on each  DIMM sub-aperture.}
They  should not  be  confused  with the  aberrations  of the  feeding
telescope.  For  example,  a  coma  or  spherical  aberration  in  the
telescope     causes     mostly     astigmatism    in     the     DIMM
sub-apertures.  Generally, the aberrations  of both  sub-apertures are
not equal, but, to simplify, we assume here their equality.

%------------------------------------------------------------------------
\subsection{Propagation effects in DIMM}
\label{sec:DIMMprop}

\begin{figure}
\includegraphics[width=8cm]{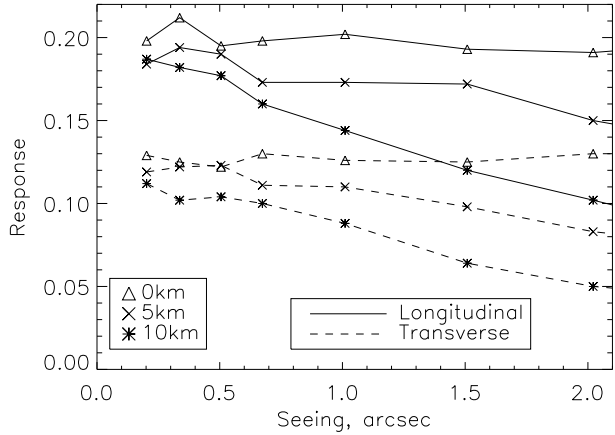} 
\caption{Effect of propagation in a DIMM with $D=0.1$\,m, $B=0.24$\,m,
 and  perfect   optics.   Monte-Carlo  simulations   with  $\lambda  =
 0.5$\,$\mu$m  and  a single  turbulent  layer  with different  seeing
 located at the ground, at 5\,km, and at 10\,km.
\label{fig:sat}}
\end{figure}

The   standard  DIMM   theory  \citep{Martin87,DIMM}   considers  only
near-field turbulence  and neglects  the propagation effects.  DIMM is
afected by three different phenomena related to propagation.

{\em  Diffraction:}  part of  the  small-scale  phase distortions  are
 converted  into amplitude  fluctuations (scintillation)  according to
 Eq.~\ref{eq:Wphi} and,  as a result,  the small-signal response  of a
 DIMM  slightly  decreases with  the  propagation  distance $z$.   The
 decrease   becomes   noticeable   for   $\sqrt{\lambda  z}   \ge   D$
 \citep{PASP02}, as can be seen in Figs.~\ref{fig:sat}--\ref{fig:def}
 and \ref{fig:Kz}.

\begin{figure}
\includegraphics[width=8cm]{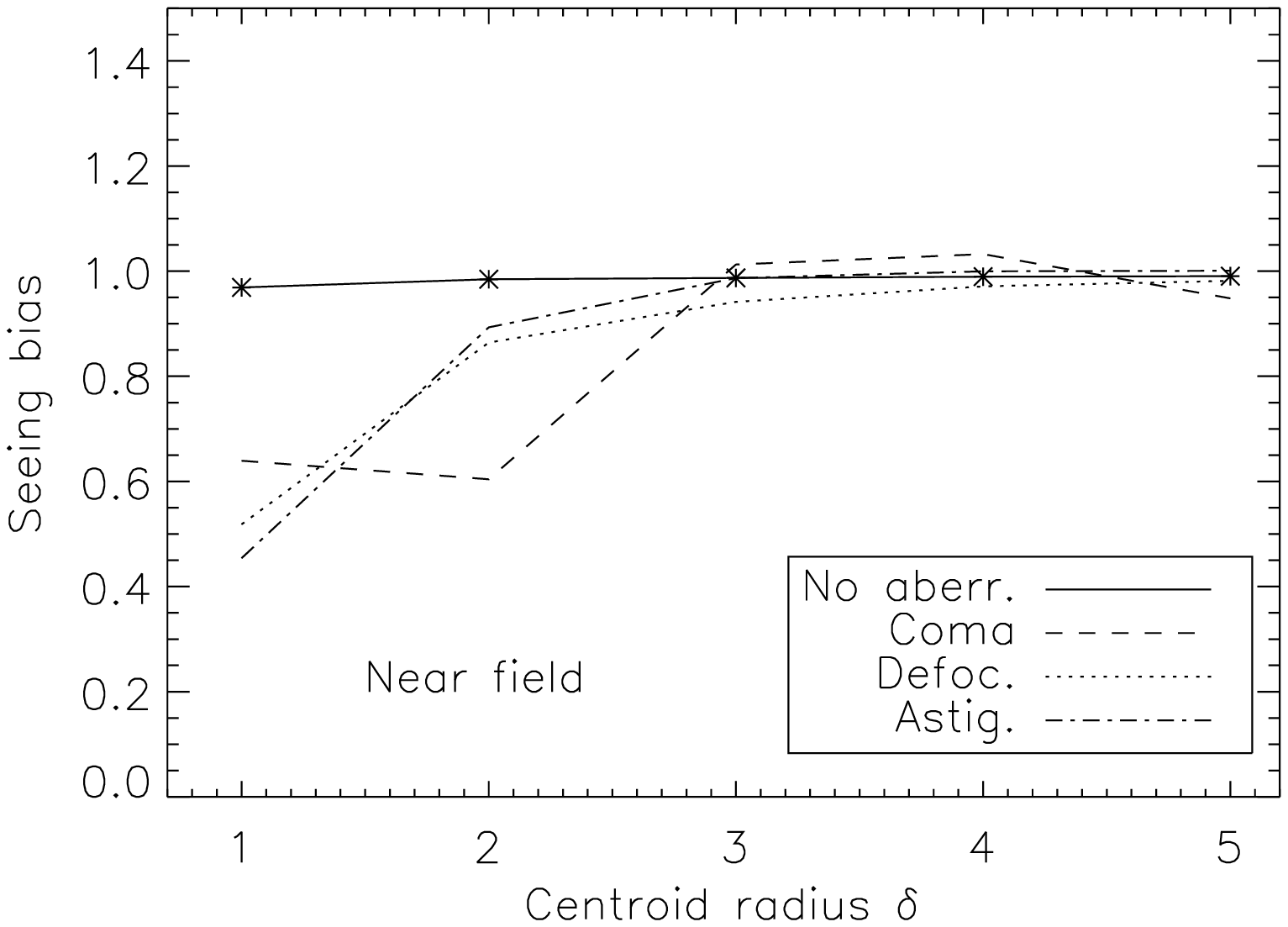} 
\includegraphics[width=8cm]{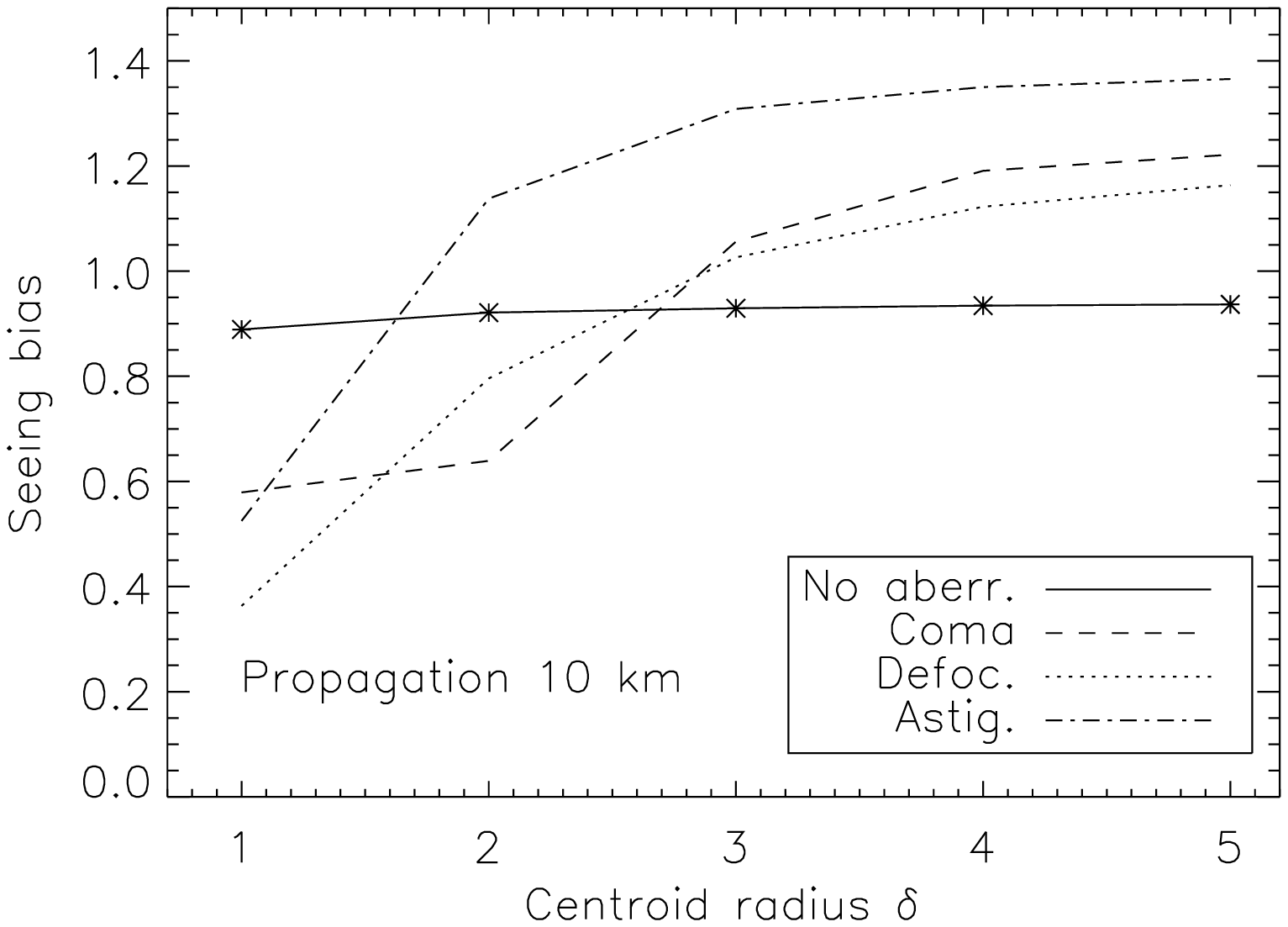}  
\caption{Bias in  the seeing  estimated from the  average longitudinal
  and transverse image  motion (Eq.~\ref{eq:seeingbias}) as a function
  of  the centroid  window radius  $\delta$ for  near field  (top) and
  10\,km   propagation  (bottom).    The  r.m.s.   amplitude   of  the
  aberrations  (coma $a_8$,  defocus $a_4$  and astigmatism  $a_5$) is
  1\,radian.  Parameters: $D=0.1$\,m, $B=0.25$\,m.
\label{fig:LT}}
\end{figure}

{\em Saturation.}  The  second, even stronger effect is  caused by the
departure from  the weak-perturbation theory and  becomes important as
soon   as  the   scintillation  index   $s^2_{\rm  D}$   exceeds  0.1.
Simulations show that the response  of a perfect DIMM to high-altitude
turbulence  is  non-linear, i.e.   the  coefficients  $K_l$ and  $K_t$
depend on the seeing (Fig.~\ref{fig:sat}).  At 0\farcs2 seeing, in the
small-signal regime, the response is reduced only by diffraction.  The
saturation causes additional loss of response, depending on the seeing
and  layer  altitude.   On  the   other  hand,  the  response  to  the
near-ground  turbulence  remains constant  even  for  a $3''$  seeing,
despite  strong distortions  of  the spots  which  split into  several
speckles.  In practise,  situations with  $s^2_{\rm D}  >0.2$  are not
uncommon and  a DIMM  is expected to  ``under-shoot''. We  could model
this effect  in the  same way  as we did  for the  MASS.  In  order to
correct for the under-shoots, we need to know the turbulence profile.

{\em  Aberrations}  together with  propagation  cause  a complex  bias
considered in the next Section.  It would be premature to correct DIMM
for  saturation   and  diffraction  before  the   aberration  bias  is
addressed.

%------------------------------------------------------------------------
\subsection{Aberrated DIMM and propagation}
\label{sec:ab-prop}

Small-signal response coefficients of an ideal and aberrated DIMMs are
calculated  with (\ref{eq:Kl})  and translated  into the  bias  in the
seeing  $\varepsilon/\varepsilon_0$ derived  from  average longitudinal
and transverse image motion,
\begin{equation}
\varepsilon/\varepsilon_0 = 
 [(K_l/K_{l,0})^{3/5} + (K_t/K_{t,0})^{3/5} ]/2,
\label{eq:seeingbias}
\end{equation}
where the  coefficients $K_{l,0}$ and  $K_{t,0}$ of an ideal  DIMM are
calculated   by    Eq.~\ref{eq:K}.    This   bias    is   plotted   in
Fig.~\ref{fig:LT} as a function of the centroid window radius $\delta$
for two  cases, turbulence  at the ground  (near field) and  at 10\,km
distance.   Both DIMM  apertures  have the  same  aberration.  In  the
near-field   case,   the  response   is   relatively  stable   against
aberrations, as long  as the centroid window is wide enough, $\delta> 2.5$.

\begin{figure}
\includegraphics[width=8cm]{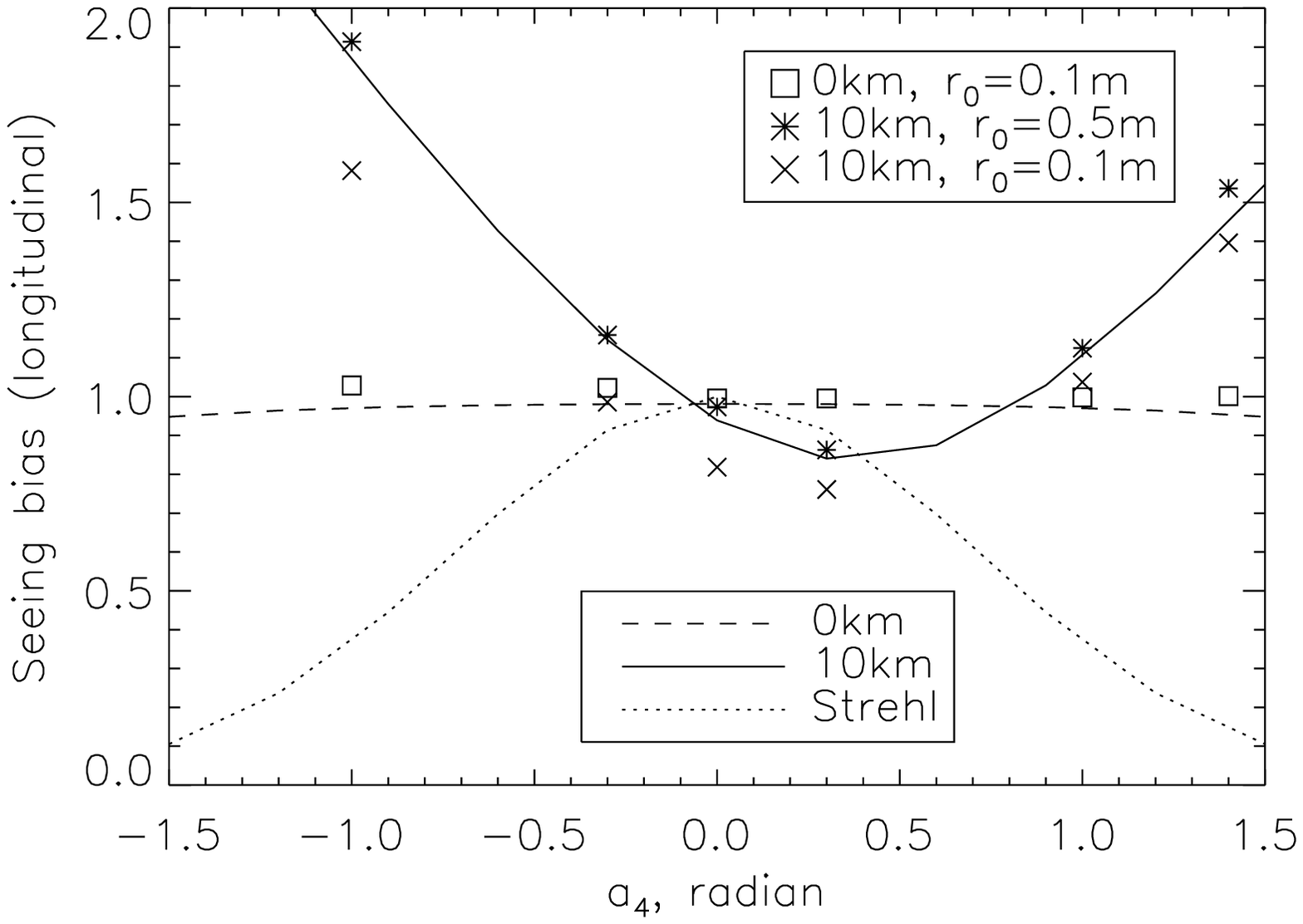} 
\includegraphics[width=8cm]{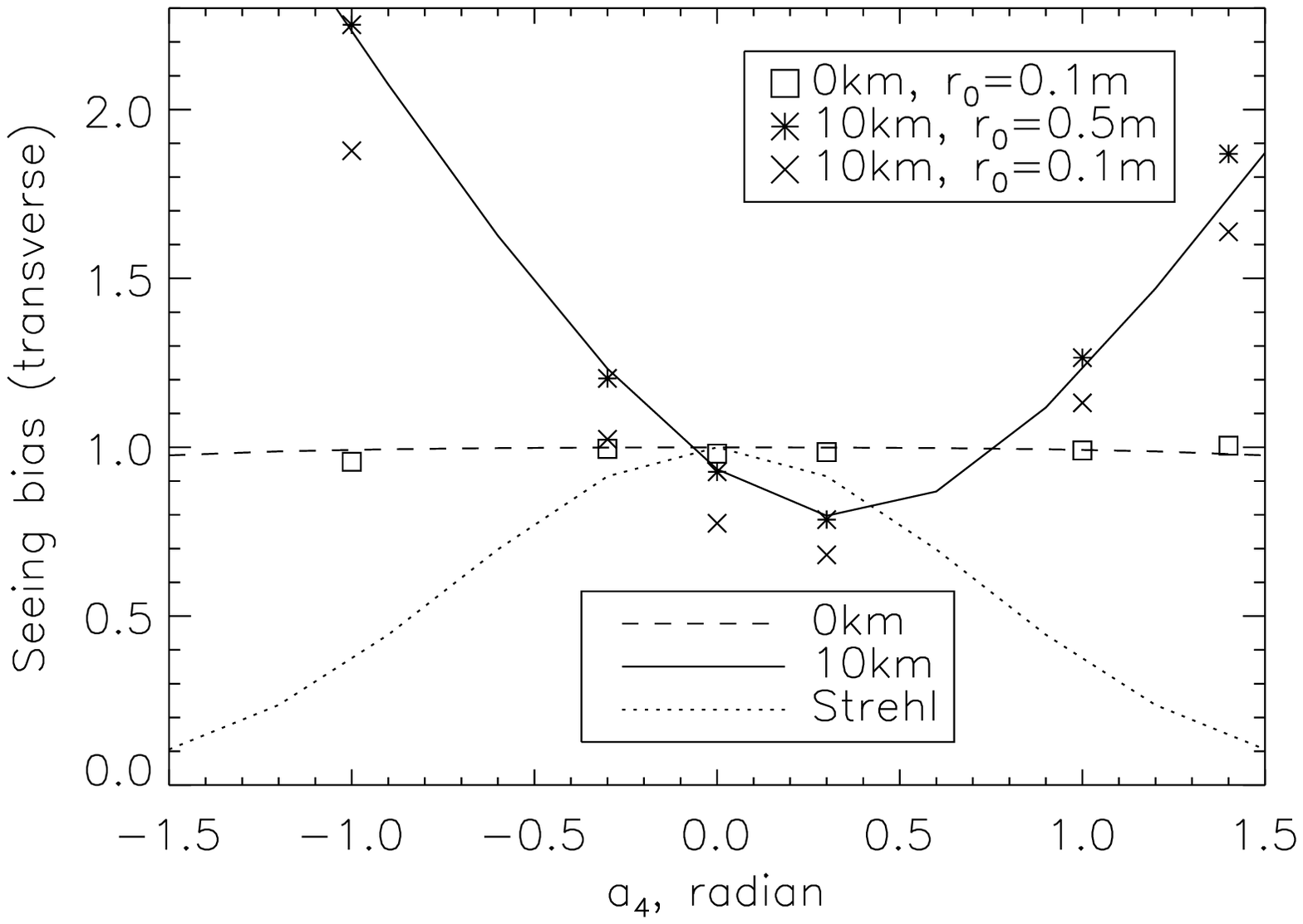} 
\caption{Bias of  the seeing derived  from the longitudinal  (top) and
transverse  (bottom)   image  motion   in  a  DIMM   with  $D=0.1$\,m,
$B=0.25$\,m,  $\delta=5$ for  turbulent layers  at the  ground (dashed
lines)  and at  10\,km  (full lines),  as  a function  of the  Zernike
defocus  coefficient $a_4$.  The  curves show  analytical calculations
for  small signal,  the results  of simulations  for  $r_0=0.1$\,m and
$r_0=0.5$\,m are over-plotted with symbols.  The dotted lines indicate
Strehl ratio.
\label{fig:def} }
\end{figure}

When the spots  are aberrated, the variance of  the differential image
motion produced by  a high turbulent layer depends on  the type of the
aberration, its amplitude, propagation distance, and the radius of the
centroid window $\delta$ (Fig.~\ref{fig:LT}).

\begin{figure}
\includegraphics[width=8cm]{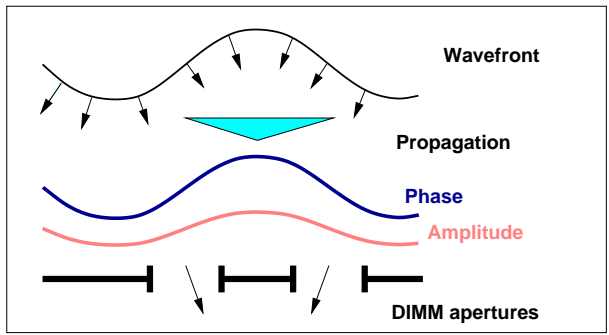} 
\caption{Illustration of the propagation effects in a DIMM (see text).  
\label{fig:propag} }
\end{figure}

The dependence  of the DIMM bias  on the defocus  amplitude is further
explored  in Fig.~\ref{fig:def}.  Even  for zero  defocus, there  is a
small difference in the  response between $z=0$ and $z=10$\,km because
of  the  diffraction.  The  difference  becomes  larger for  defocused
spots.  The influence  of a small defocus depends on  its sign: on one
side of  the focus, the response decreases  and DIMM ``under-shoots'',
on the other side it  increases. A defocus larger than 0.8\,rad always
causes a  positive bias.  These analytical  calculations are confirmed
by simulations.   When $r_0=  0.5$\,m, the simulated  response closely
follows  the small-signal  curves.  For  $r_0=0.1$\,m,  the near-field
response still matches  the theory, but the 10-km  response is reduced
additionally by the saturation (Fig.~\ref{fig:sat}).

  A negative  bias caused  by the  combination of  propagation and
small   defocus   may  appear   counter-intuitive:   we  expect   that
scintillation  adds something  to  the image  motion  produced by  the
phase.   Figure~\ref{fig:propag}  illustrates  this  effect  from  the
geometric-optics perspective.  Suppose  that longitudinal image motion
in a DIMM is created by  a sinusoidal wavefront with a period $\approx
2B$.  After propagation, the phase  is reduced only slightly, but some
amplitude fluctuations  are created, positive where  the rays converge
and   negative  where   they  diverge   (amplitude   fluctuations  are
proportional  to  the  wave-front  curvature).  The  gradient  of  the
amplitude  at DIMM  apertures is  correlated with  the  phase gradient
(i.e.  image motion).  A perfect  DIMM is insensitive to the amplitude
fluctuations, but  in a defocused  DIMM the centroids are  affected by
the amplitude  gradients.  Depending on  the sign of the  defocus, the
amplitude gradient  either increases  or decreases the  measured image
motion and thus creates a positive or negative bias.  

A positive bias  caused by strong defocus (and  other aberrations) has
been   confirmed   experimentally   by  direct   comparisons   between
well-aligned  and aberrated DIMMs  \citep{TMT-DIMM}. The  negative bias
found here has  not been suspected before.  A  defocus of 0.3\,rad rms
(24\,nm for $\lambda = 0.5$\,$\mu$m)  corresponds to a Strehl ratio of
0.91, i.e.   practically diffraction-limited spots.  Yet,  such a DIMM
can have response coefficients reduced or increased by $\sim$20\%, and
will produce  seeing data  biased by $\sim  \pm12$\% when most  of the
turbulence is concentrated at  10\,km.  Optical aberrations other than
defocus are  expected to cause  a similarly complex bias.   We conclude
that  {\em the  response  of  a DIMM  to  high-altitude turbulence  is
intrinsically  inaccurate.}  The saturation  makes things  even worse,
introducing dependence of the response on turbulence profile.

We   simulated  the   centroid  calculation   by   {\em  thresholding}
numerically  and found  that  its  behaviour is  very  similar to  the
windowing method  considered above. For turbulence at  the ground, the
response  is insensitive to  aberration, but  slightly depends  on the
threshold (a  bias of $+10$\%  in $K_l$ and  $K_t$ for a  threshold of
0.1).   The   asymmetric  dependence  of   response  to  high-altitude
turbulence on the defocus (Fig.~\ref{fig:def}) is also preserved.

\begin{figure}
\includegraphics[width=8cm]{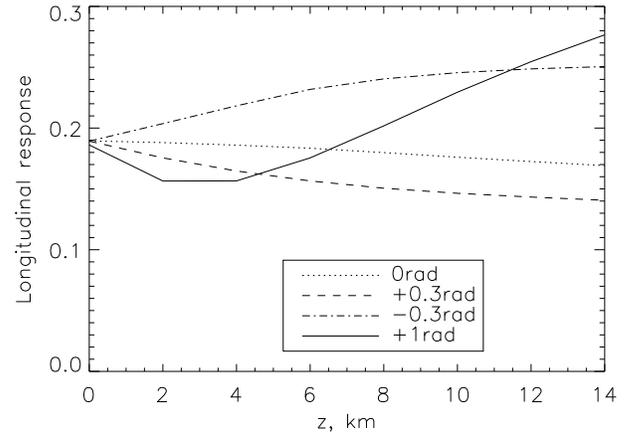} 
\caption{The dependence of  longitudinal 
small-signal  response   coefficient $K_l$  of  a   DIMM  with  $D=0.1$\,m,
$B=0.25$\,m,  $\delta=5$ on  the  propagation distance  $z$ for four
values of defocus indicated on the plot. The behaviour of the
transverse coefficient is similar. 
\label{fig:Kz} }
\end{figure}

Figure~\ref{fig:Kz} shows how the small-signal response depends on the
propagation distance $z$. For  un-aberrated spots, the response slowly
decreases  with $z$.   On the  other hand,  defocused spots  show both
positive  and  negative  bias.   A  real DIMM  instrument  can  either
``over-shoot''  or  ``under-shoot'', depending  on  the  state of  its
optics and the turbulence profile.

It is clear  that the quality of  the DIMM optics and focus  has to be
strictly controlled  in order to get unbiased  results. Measurement of
the spot Strehl ratio (SR) suggested in \citep{PASP02} is helpful, with
SR$>0.5$   usually  indicating  acceptable   quality  \citep{TMT-DIMM}.
However, the SR  is reduced under poor seeing even  in a perfect DIMM,
so if the data with poor  SRs are rejected, the seeing statistics will
be biased.  Yet another way to quantify the aberrations in a DIMM will
be to take long exposures of  defocused spots and to analyse them with
the  method described  by \citet{Donut}.  Needless to  say that  a DIMM
working always out-of-focus    \cite[e.g.][]{Bally96} will never
measure the  seeing accurately.

%---------------------------------------------------------------
\subsection{Exposure-time bias}

Finite exposure time in a  DIMM reduces the differential image motion,
biasing  the measured  seeing to  smaller values.   The effect  can be
quite strong, as noted by \citet{Martin87}.  This bias was modelled in
detail  in \citep{PASP02}.   Short (e.g.  5\,ms) exposures  reduce the
bias, but  do not  eliminate it completely.   If the data  is acquired
continuously, an  extrapolation to zero-exposure can be  done from the
variances  of time-binned  signal,  either by  fitting an  exponential
curve   to  the   dependence  of   the  variance   on   exposure  time
\citep{TMT-DIMM} or by using a simple linear formula as in MASS or GSM
\citep{GSM}.  The linear extrapolation appears too strong, though.

If the image sequence is not continuous, a method of interlaced single
and  double exposures  should be  used.   In this  case, a  ``modified
exponential        correction''         was        developed        in
\citep{PASP02}\footnote{Formula   13   in   \citep{PASP02}  contains   a
typographic   error,   corrected   here}.   If   $\varepsilon_1$   and
$\varepsilon_2$  are the  seeing  values calculated  with nominal  and
double  exposure   time,  the  de-biased   seeing  $\varepsilon_0$  is
estimated from
\begin{equation}
\varepsilon_0 \approx 0.5 (c_1 \varepsilon_1 + c_1^{7/3} \varepsilon_2),
\label{eq:exp-bias}
\end{equation}
where  $c_1  =  (\varepsilon_1  /  \varepsilon_2)^{3/4}$.   To  reduce
the statistical noise, the factor $c_1$ is averaged (smoothed)  over time and
its  average value  is  then used  in  (\ref{eq:exp-bias}) to  correct
 individual measurements.

%------------------------------------------------------------------------
\subsection{Centroid noise}

Even  in  the  absence  of  atmospheric  image  motion,  the  measured
centroids fluctuate because  of the errors caused by  the photon noise
and detector  readout noise.  The errors of  intensities $I_{i,j}$ are
independent in each pixel and equal  to the sum of readout and Poisson
noise, expressed in the signal counts (ADU):
\begin{equation}
\sigma^2_{I_{i,j}} = R^2 + I_{i,j}/G,
\end{equation}
were  $R$  is  readout noise  in  ADU,  and  $G$  --- the  CCD  camera
conversion factor (gain) in  $e^-$/ADU.  The influence of these errors
on  the calculated  centroid  is easily  evaluated by  differentiating
Eq.~\ref{eq:centr} and using independence  of noise in each pixel.  As
the weights $w_{i,j}$ take values of either 1 or 0, we simply restrict
the summation to pixels where $w_{i,j} = 1$. The result is
\begin{eqnarray}
\sigma^2_c  & = & \frac{1}{I_0^2} \sum_{i,j} (x_{i,j} - c_x)^2 \sigma^2_{I_{i,j}}
\nonumber \\
& = & \frac{1}{I_0^2} \sum_{i,j} (x_{i,j} - c_x)^2 (R^2 + I_{i,j}/G)
\label{eq:sig3}
\end{eqnarray}
Here  $I_0$ is  the sum of intensities over pixels in the centroid
window.

The sum entering in (\ref{eq:sig3}) can be computed in advance if the
centroid  window has  a well-defined  size  and the  image profile  is
known.   This is  not the  case when  a thresholding  method  is used.
However, even with thresholding  the centroid noise of each individual
spot   can   be  evaluated   with   (\ref{eq:sig3})  during   centroid
computation.  Variations of the flux caused by scintillation or clouds
can be accounted for as well.

Obviously, the noise variance of both  centroids in the DIMM has to be
evaluated  and  subtracted  from  the measured  differential  variance
before  calculating the  seeing  with (\ref{eq:KDIMM}).   It would  be
wrong to  express the noise in  arc-seconds and subtract  it later from
the  measured seeing,  because  the effects  in  the longitudinal  and
transverse  directions are  not the  same.  The  noise depends  on the
stellar flux,  hence subtracting a  fixed ``instrumental
noise''  is  not correct.   Typically,  the  noise  is small  and  its
subtraction  or   not  does  not  matter.   However,   the  noise  can
significantly bias DIMM results under  good seeing or for faint stars.
The flux in  the interlaced single and double  exposures is different,
hence if the  noise variance is not properly  subtracted, it will bias
(over-estimate) the correction (\ref{eq:exp-bias}).

To reduce the  noise, especially the term related  to $R$, the smallest
number of pixels must be  used, favouring a narrow centroid window with
$\delta \sim 1$ or, equivalently,  a high threshold.  However, we have
seen  in Sect.~\ref{sec:resp-real} that  such choice  leads to  a very
uncertain   response   coefficient   in   presence   of   even   small
aberrations.  The noise  is usually  much less  of a  problem  than the
aberrations, so a wide centroid window  with $\delta \ge 2.5$ or a low
threshold  are preferable,  contrary  to the  recommendation given  in
\citep{PASP02}.

Both the measurement  noise and the noise caused  by the scintillation
are  isotropic and  affect  longitudinal and  transverse image  motion
equally.   The  effect on  the  calculated  seeing,  however, will  be
different  because  $K_l  >  K_t$.   If the  DIMM  signal  contains  a
significant noise component, the  ``transverse'' seeing will always be
larger  than  the  ``longitudinal''  seeing.   This  effect  could  be
mistakenly  interpreted   as  a  manifestation   of  a  non-Kolmogorov
turbulence spectrum.

%------------------------------------------------------------------------
%------------------------------------------------------------------------
%------------------------------------------------------------------------

%------------------------------------------------------------------------
\section{Conclusions}
\label{sec:concl}

This  work complements  previous  studies of  two  methods to  measure
seeing --  DIMM and  MASS -- and  focuses on the  achievable accuracy.
The  ``seeing'' itself  cannot  be defined  very  accurately, being  a
model-dependent  parameter  of a  random  and non-stationary  process.
Taking aside this caveat, we investigate potential instrumental biases
by  simulating both  turbulence and  the instruments  numerically. The
true  seeing is  known  exactly, permitting  to calibrate  the
methods on the absolute scale.

Our conclusions and recommendations can be summarised as follows:
\begin{itemize}

\item
The  departure  from the  weak-perturbation  theory  affects the  MASS
method  seriously,   but  it  is  correctable   for  a  not-too-strong
scintillation, $s_{\rm A}^2 < 0.7$.

\item
The effects of finite exposure time and inner turbulence scale in MASS
can be neglected.

\item
The DIMM  method is  very robust and  tolerant to aberrations  for the
near-ground turbulence, provided that  the centroid calculation uses a
large enough radius, $ > 2.5 \lambda/D $.

\item
Optical propagation causes a non-linear response of DIMM (saturation),
 not present in the near-field case.

\item
Combination  of propagation  and  aberrations  in a  DIMM  leads to  a
complex bias.   By controlling the DIMM optical  quality (Strehl ratio
$>0.5$), we can keep the seeing  bias to within $\pm 12$\%, but making
the control tighter appears impractical.

\item
Centroid noise  in a DIMM should  be computed and  subtracted from the
measured variance. To do so, the detector readout noise and gain must be known.

\item
The  exposure-time bias in  a DIMM  should be  corrected using  one of
several known recipes.

\end{itemize}

The sensitivity  of the DIMM  to propagation  and aberrations
comes as  a surprise, although  neither of these effects  was accounted
for by the standard, near-field DIMM theory. The bias on high-altitude
seeing is so complex that removing it completely seems unrealistic. 

It has been demonstrated that two identical DIMM instruments with good
optics can  give seeing measurements concordant to  within few percent
\citep{TMT-DIMM}.  Are  these measurements  {\em accurate} to  the same
level? Not necessarily. A seeing  of $1''$ coming from 10\,km could be
measured by both instruments with a bias of 0.83 (Fig.~\ref{fig:sat}),
whereas the same  seeing originating near the ground  will be measured
correctly.

The  response of  a  DIMM  depends on  both  instrumental factors  and
observing  conditions. Two  different  DIMMs can  agree  on one  night
(e.g. seeing dominated by low layers) and disagree on another night or
at another site. Two identical DIMMs  can be biased in a different way
at two different sites.  Thus, inter-comparison between DIMM (or MASS)
instruments  cannot  guarantee that  they  are  {\em accurate}.  Their
mutual agreement is a necessary,  but not sufficient condition. Only a
careful control  of biases can ensure accurate  seeing data.  However,
instrument inter-comparisons  are  useful for debugging  and checks
and should be pursued whenever possible.

A combination of  MASS and DIMM in one  instrument has stimulated this
research.   These instruments,  when properly  calibrated,  agree very
well for  a seeing dominated by  high layers \citep{Kor07}.   This is a
triumph  of the optical  propagation theory  enabling us  to interpret
both scintillation and  image motion with a common  model and a single
parameter,  $r_0$.   At  the  same  time, the  agreement  between  two
instruments based  on different principles and  with different biases
is a  strong argument that both  are {\em accurate},  i.e. measure the
seeing on the absolute scale.

%----------------------------------------------------------------------
\section*{Acknowledgements}

 The  development  of the  MASS-DIMM  instrument  and  the methods  of
getting  accurate seeing data  has been  stimulated and  encouraged by
many people  and organisations. We  acknowledge the support  from NOAO
and  ESO in  building and  testing  the initial  prototypes and  final
instruments.  Several testing campaigns have  been done at CTIO in the
period 2002-2004.
The software of  the MASS instrument has been  developed and supported
by  the   team  of  young   talented  astronomers  at   the  Sternberg
Astronomical  Institute  of   the  Moscow  University  --  N.~Shatsky,
O.~Voziakova, S.~Potanin.  The mechanical  parts were produced  at the
CTIO  Workshop.    We  are  indebted  to  the   CTIO  ``sites  group''
(E.~Bustos,  J.~Seguel,  D.~Walker)  for  maintaining  MASS-DIMM  site
monitors in a working and well-calibrated condition.

%------------------------------------------------------------------------

%------------------------------------------------------------------------
\appendix

%------------------------------------------------------------------------
\section{Response of a centroid estimator in the pupil plane}
\label{sec:PResp}

The  centroid signal  $c$  for each  DIMM  spot is  obtained from  the
weighted PSF $I({\bfit a})$ as

\begin{equation}
c = I_0^{-1} \; \int {\rm d}^2 {\bfit a}\; I({\bfit a}) \; M({\bfit a})   ,
\label{eq:c}
\end{equation}
where  ${\bfit a}$  is a  2-dimensional vector  of  angular coordinates,
$M({\bfit  a})$ is some  function, {\em  mask}, and  $I_0$ is  the total
intensity  (flux).  All  integrals are  in infinite  limits  and exist
because all functions  are supposed to have limited  support.  The PSF
is not  necessarily an  ideal one, but  may include  some aberrations.
Formula (\ref{eq:c}) is rather general and applies to many situations,
e.g.  curvature sensing.   For centroid  calculation, we  need  a mask
$M({\bfit a})  = a_x  w({\bfit a})$ to  match Eq.~\ref{eq:centr}.  We keep
(\ref{eq:c}) in a general form useful for other applications.

Let  ${\bfit x}$  be  the coordinate  vector  in the  pupil plane.   The
complex amplitude  of the initial  un-perturbed field at the  pupil is
$U({\bfit  x})$.    It  includes  the  pupil   function  (possibly  with
aberrations)  and   is  normalised  arbitrarily.    Suppose  that  the
amplitude $U$  is changed by  a small phase aberration  $\varphi ({\bf
x})$  and a  small  log-amplitude perturbation  $\chi  ({\bfit x})$  and
becomes $ U({\bfit x})  \; e^{ i \varphi ({\bfit x}) +  \chi ({\bfit x})} $.
What  would be  the change  of the  signal $\Delta  c$ caused  by this
aberration?

We  find  a  small change  in  the  signal  by linearising  the  known
expression of the OTF  \citep{Goodman} 
\begin{equation}
\tilde{I}({\bfit f}) = I_0^{-1} \int {\rm d}^2 {\bfit x} \; U({\bfit x}) U^*({\bfit x} + \lambda {\bfit f}) 
\label{eq:OTF}
\end{equation}
with respect to  small perturbations $\varphi \ll 1$  and $\chi \ll 1$,
so-called {\em PSF Taylor expansion} \citep{Perrin}:
\begin{eqnarray}
\Delta \tilde{I}({\bfit f}) &  = & I_0^{-1} \int {\rm d}^2 {\bfit x} U({\bfit x}) 
U^*({\bfit x} + \lambda {\bfit  f}) \label{eq:dOTF} \\
& \times & [ i \varphi({\bfit x}) + \chi({\bfit x}) -   i \varphi({\bfit x} +
  \lambda {\bfit f} ) +  \chi({\bfit x} + \lambda {\bfit f}) ] \nonumber
\end{eqnarray}

The signal increment is 
\begin{equation}
\Delta c = \int {\rm d}^2 {\bfit f} \;   \Delta \tilde{I}({\bfit f}) \tilde{M}( {\bfit f}) .
\label{eq:deltac}
\end{equation}
We  put  (\ref{eq:dOTF})   into  (\ref{eq:deltac})  and  re-group  the
terms.  The  first  term  in   the  square  brackets  containing  $i
\varphi({\bfit x})$ leads to
\begin{eqnarray}
\Delta c_1 & = & i (\lambda^2 I_0)^{-1} \; \int {\rm d}^2 {\bfit
 x}   \int   {\rm d}^2 {\bfit x'} \; 
 U({\bfit x})  U^*({\bfit x} +  {\bfit  x}')  \nonumber \\
& \times & \tilde{M}( {\bfit x}'/\lambda)  \; \varphi({\bfit x})   ,
\label{eq:deltac2}
\end{eqnarray}
where ${\bfit x}'  = \lambda {\bfit f}  $. The 3-rd term containing  $ - i
  \varphi({\bfit x} + \lambda {\bfit  f} )$ leads to the complex-conjugate
  of the  expression (\ref{eq:deltac2}) with  inverse sign. Collecting
  all 4 terms, we write the result as
 \begin{equation}
\Delta c =   \int  {\rm d}^2
       {\bfit x} \; F_{\varphi} ({\bfit x}) \; \varphi ({\bfit x}) \;  +   
 \int  {\rm d}^2 {\bfit x} \; F_{\chi} ({\bfit x}) \; \chi ({\bfit x})  , 
\label{eq:deltac1}
\end{equation}
where the {\em filter functions} $F$ are 
\begin{equation}
 F_{\varphi} ({\bfit x}) = 
  {\rm Im} [ A({\bfit x}) ], \;\;\;\;
 F_{\chi} ({\bfit x}) =  {\rm Re} [ A({\bfit x}) ] 
\label{eq:F}
\end{equation}
and 
\begin{eqnarray}
A({\bfit x}) =   2 \; (\lambda^2 I_0)^{-1} \; U({\bfit x})
\int  {\rm d}^2
      {\bfit x'}  \; U^*({\bfit x} + {\bfit x'}) \; \tilde{M}({\bfit x'} /\lambda) 
%\nonumber \\
%=  2 ( \lambda I_0)^{-1} \; U \; [ U^* \star \tilde{M} ] .
\label{eq:A}
\end{eqnarray}
%
%In the symbolic notation  $\star$ denotes correlation and the argument
%scaling  of $\tilde{M}$ (Fourier  transform of  $M$) is  omitted.  
The response is independent of  the normalisation of the amplitude $U$
because it is divided by the flux.

This result  contains an implicit assumption that  the fluctuations of
the denominator $I_0$  can be neglected. This is  not always true. The
signal has the form $c =  A/B$, hence its fluctuations are $\Delta c =
\Delta  A/B  -  c   (\Delta B/B)$.   The  fluctuations  of  the
denominator $\Delta  B$ can be  neglected if the average  signal $c=0$
(they will be a second-order term then). This condition is enforced by
the choice of $M({\bfit a}) = a_x w({\bfit a})$ appropriate for a windowed
centroid. 

If the mask  $M$ is correctly dimensioned to  compute the centroid $c$
in pixels  and the angular size of  the pixel is $p$,  then it follows
that the longitudinal response coefficient $K_l$ is
\begin{eqnarray}
K_l  & = &  ( p  D/\lambda)^2 \;  (r_0/D)^{5/3} \;  \int {\rm d}^2 {\bfit f} \nonumber \\
& \times & |\tilde{F}_\varphi ({\bfit f})|^2 \; \Phi_\varphi({\bfit f}) [ 2 \sin (\pi
  B f_x) ]^2 ,
\label{eq:Kl}
\end{eqnarray}
where both centroid measurement direction and baseline are parallel to
the $x$ axis.  To calculate  the transverse response $K_t$, we replace
$f_x$  with  $f_y$.   To   take  into  account  both  propagation  and
sensitivity to scintillation, we make a replacement
\begin{equation}
  |\tilde{F}_\varphi({\bfit f})|^2 \rightarrow
  | \tilde{F}_{\varphi}({\bfit f}) \cos (\pi \lambda z |{\bfit f}|^2)
 +  \tilde{F}_{\chi}({\bfit f}) \sin (\pi \lambda z |{\bfit f}|^2) |^2 .
 \label{eq:fresnel}
\end{equation}
This modification  automatically accounts for  the correlation between
phase and  amplitude.

%------------------------------------------------------------------------
%------------------------------------------------------------------------

\bsp

\label{lastpage}

\end{document}